\documentclass{article}

\usepackage{arxiv}

\usepackage[utf8]{inputenc} % allow utf-8 input
\usepackage[T1]{fontenc}    % use 8-bit T1 fonts
\usepackage[hidelinks]{hyperref}       % hyperlinks
\usepackage{url}            % simple URL typesetting
\usepackage{amssymb}

\usepackage{authblk}
\usepackage{booktabs}       % professional-quality tables

\usepackage{amsfonts}       % blackboard math symbols
\usepackage{nicefrac}       % compact symbols for 1/2, etc.
\usepackage{graphicx}
\usepackage{doi}
\usepackage{float}
\usepackage{stfloats}
\usepackage{subcaption}
\usepackage{textcomp, gensymb}
\usepackage{enumitem}
\usepackage{multirow}
\usepackage{xcolor}
\usepackage{pgfplots}
\usepackage{wrapfig}
\usepackage{rotating}
\usepackage{todonotes}
\usepackage[ruled]{algorithm2e}
\usepackage{amsmath}
\usepackage{bbm} % \mathbbm{1}
\usepackage[most]{tcolorbox}
\usepackage{xparse}
\usepackage{enumitem}
\usepackage{array}
\newcolumntype{x}[1]{>{\centering\arraybackslash\hspace{0pt}}p{#1}}

\usepackage{colortbl}
\definecolor{posgreen}{RGB}{198, 239, 206}
\definecolor{negred}{RGB}{255, 199, 206}

\usepackage{amsthm}
\newtheorem{proposition}{Proposition}

\pgfplotsset{compat=1.18}

\usepackage{tikz}
\usepackage{tikz-qtree}
\usetikzlibrary{trees}
\usetikzlibrary{mindmap}
\usetikzlibrary{positioning}

\theoremstyle{plain}
\newtheorem{assumption}{Assumption}
\theoremstyle{remark}
\newtheorem{remark}{Remark}

\newcounter{llmprompt}

\NewEnviron{llmprompt}[3][0.9\linewidth]{%
  \refstepcounter{llmprompt}%
  \def\prompttitle{Prompt~\thellmprompt}%
  \if\relax\detokenize{#2}\relax
  \else
    \edef\prompttitle{Prompt~\thellmprompt: #2}%
  \fi
  \begin{tcolorbox}[float=!ht,
    width=#1,
    colback=gray!10,
    colframe=black,
    fonttitle=\bfseries,
    title=\prompttitle,
    sharp corners,
    boxrule=0.5pt,
    arc=2mm,
    left=4pt,
    right=4pt,
    top=4pt,
    bottom=4pt,
    listing only,
    listing options={
      basicstyle=\ttfamily\small,
      breaklines=false,
      breakautoindent=false,
      columns=fullflexible,
      keepspaces=true,
      showstringspaces=false,
      aboveskip=0pt,
      belowskip=0pt,
      frame=none,
      rulecolor=\color{white},
      rulesepcolor=\color{white}
    },
    label={#3}
  ]
  \BODY
  \end{tcolorbox}
}

\newtcolorbox{intuitionbox}{
  colback=blue!4,
  colframe=blue!45!black,
  fonttitle=\bfseries,
  title=Intuition,
  boxrule=0.5pt,
  sharp corners,
  enhanced
}

\tcbset{
  myresultbox/.style={
    colback=white,       % no background color
    colframe=gray,      % black frame
    fonttitle=\bfseries, % bold title
    title=Result~\thetcbcounter,
    boxrule=0.5pt,       % thin line
    sharp corners,
    enhanced,
  }
}

\newtcolorbox[auto counter, number within=section]{resultbox}[1][]{
  myresultbox, #1
}

\tcbset{
  myobservationbox/.style={
    colback=white,       % no background color
    colframe=gray,      % black frame
    fonttitle=\bfseries, % bold title
    title=Observation~\thetcbcounter,
    boxrule=0.5pt,       % thin line
    sharp corners,
    enhanced,
  }
}

\newtcolorbox[auto counter, number within=section]{observationbox}[1][]{
  myobservationbox, #1
}

\newcounter{rqcounter}
\newcommand{\RQ}[1]{%
    \stepcounter{rqcounter}%
    \item{\textbf{RQ\therqcounter}: #1}%
}

\begin{document}

\title{Label-Free Detection of Structural Bias in LLM-Based Recommender Systems via Randomized Paraphrasing: The Case of Verbosity}
\newcommand{\shorttitle}{Label-Free Structural Bias detection in LLM-Based RS: Verbosity as a Case Study}

\author{Carlos Pel\'aez-Gonz\'alez$^{1,2,*}$, David Herrera-Poyatos$^{1,2}$, Francisco Herrera$^{1,2}$, Andr\'es Herrera-Poyatos$^{3}$}

\affil{$^{1}$Department of Computer Science and Artificial Intelligence (DECSAI), University of Granada, 18014 Granada, Spain. \\
$^{2}$Andalusian Research Institute in Data Science and Computational Intelligence (DaSCI), 18016 Granada, Spain. \\
$^{3}$Department of Algebra, Faculty of Sciences, University of Granada, 18071 Granada, Spain. \\
$^{*}$Corresponding author. \\
Emails: \texttt{\{carlosprog, divadhp, andreshp\}@ugr.es}, \texttt{herrera@decsai.ugr.es}}

\date{\today}

\maketitle              % typeset the header of the contribution

\begin{abstract}
Large language models (LLMs) are increasingly deployed as recommenders in e-commerce and conversational shopping systems, where they select a product from a set of retrieved candidates described in natural language. Their decisions should depend on relevance to the user query, but may also be influenced by surface-level properties of the descriptions, such as their length. Detecting such structural biases is difficult because relevance labels are rarely available. We propose a label-free framework to audit LLM-based recommender systems for structural bias. Candidate descriptions are rewritten by randomly assigned paraphrasing models and presented in random order. We prove that, if the recommender's choice is invariant to these rewritings, every paraphraser is selected with equal probability, yielding a statistical test that requires no relevance labels. A four-step statistical pipeline tests this property and quantifies the contribution of each description feature while controlling for query--product relevance. We apply the framework to verbosity bias on 20{,}000 product queries and nine recommender LLMs from five model families. After controlling for the measured relevance signals, description length has a positive effect on selection in all nine models, increasing the odds of selection by up to 38\% per standard deviation of length. Relevance remains the primary driver of recommendations, while verbosity acts as a consistent secondary signal whose strength depends on the model family. Since the framework needs only black-box access and no annotations, it can be run as a routine audit of deployed shopping assistants, and we provide an auditing protocol with cost and sample-size guidance. The source code is publicly available\footnote{\url{https://github.com/GoComputing/Verbose-Bias.git}}.
\end{abstract}

\keywords {Large language models \and LLM-based recommender systems \and structural bias \and verbosity bias}

\section{Introduction}

% Brief introduction to LLMs and zero-shot capabilities. Then introduce the concept of RS and LLM-based RS
Large language models (LLMs) have emerged as powerful general-purpose reasoners, demonstrating strong zero-shot and few-shot performance across a broad spectrum of tasks, from question answering and code generation to text summarization and machine translation, without requiring task-specific training~\cite{zhao2026surveylargelanguagemodels}. Their ability to process and generate natural language has made them natural candidates for integration into recommender systems~(RS)~\cite{lin2025how}, traditionally built on collaborative filtering or content-based methods that model user--item interactions through learned embeddings~\cite{10.1145/3285029, BOBADILLA2013109}. LLM-based RS leverage this linguistic competence to represent user preferences and items directly in natural language, enabling rich, flexible descriptions of preferences and products within a single prompting interface~\cite{li-etal-2024-large, zhao2024recommender}. Among the paradigms explored for LLM-based RS, list-wise ranking, where the model receives a set of candidate items and directly outputs a ranked ordering or a single selected item, has attracted particular attention for its simplicity and its natural fit with instruction-following LLMs~\cite{hou2024llmrank}. This paradigm, however, introduces a structural vulnerability: by presenting all candidates simultaneously as plain text, it exposes the model to systematic sensitivities to surface-level properties of item descriptions.

% Different types of RS. We focus on top-1, single interaction RS
Recommender systems differ in how many items they surface and how they interact with the user. In conversational recommender systems, which elicit preferences through multi-turn dialogue, LLMs now play three distinct roles: as the recommendation engine itself, as supporting components that augment specific parts of the pipeline (e.g.\ data augmentation, re-ranking or grounding), and as tools for user simulation and evaluation~\cite{KIM2026132508}. Top-1 and single turn LLM-based RS, the focus of this work, use the LLM in the first of these roles but in a single turn: they receive a set of $k$ candidate items retrieved from a large catalog and select a single recommended product in response to a natural language query, a paradigm that directly mirrors high-stakes real-world deployments such as e-commerce product search, voice-driven shopping assistants, and conversational product discovery interfaces~\cite{li-etal-2024-large}. In these settings, the model commits to a single recommendation with no opportunity for iterative preference refinement, making the quality of that single decision entirely dependent on how the model processes the candidate descriptions presented in its input prompt. This also makes the setting particularly consequential: unlike multi-turn systems where an erroneous first recommendation can be corrected through dialogue~\cite{KIM2026132508}, a systematic flaw in how the model reads those descriptions significantly degrades the user experience. Understanding what drives the recommender's choice among candidates and whether its decisions reflect only query-product relevance or are also shaped by other properties of the descriptions is therefore a question of direct practical importance.

This question has a concrete operational counterpart. Organizations deploying LLM-based shopping assistants routinely replace the underlying model, update the prompt or ingest new catalog content, and each of these changes may silently alter which products are recommended. Auditing such changes with standard offline evaluation requires relevance judgments for every query--product pair, which are expensive to collect, rarely available for live catalogs, and quickly outdated. In marketplaces where product descriptions are written by third-party sellers, an undetected sensitivity to how descriptions are written also distorts competition between sellers and creates incentives to optimize listings for the recommender rather than for customers. Platform operators therefore need audit procedures that work with black-box access to the deployed system, on its own catalog and queries, and without annotation effort, a need that is reinforced by emerging regulation requiring transparency and risk assessment of recommender systems on large online platforms~\cite{dsa2022}.

% Introduction to structural biases and verbosity bias. Thus, we wonder about the following questions
% A natural concern in this setting is whether the recommender's decisions truly reflect query-product relevance, or whether they are also systematically shaped by properties of the candidate descriptions that are orthogonal to their content. 
Prior work has documented that LLMs acting as evaluators or rankers exhibit a range of \emph{structural biases}: systematic tendencies to favor candidates based on where they appear in the prompt, how they are worded, or how long their descriptions are, independently of their actual quality~\cite{GU2026101253}. In the recommendation setting, where the model selects among products described in natural language, these biases are particularly consequential: a model that systematically favors longer or more elaborate descriptions may consistently recommend products that are not the most relevant, simply because their descriptions happen to be more verbose. Despite growing evidence of such effects in LLM-as-judge settings, their presence and magnitude in LLM-based RS remain largely uncharacterized. In particular, \emph{verbosity bias}, the tendency to assign higher scores to longer or more elaborate inputs independently of their relevance, has received considerably less rigorous treatment than positional bias, and no statistically grounded methodology exists to detect it in a recommendation context without access to ground-truth relevance labels. This motivates the following research questions:

\begin{itemize}
    \RQ{Do LLM-based RS exhibit verbosity bias in product selection?}
    \RQ{Does verbosity independently influence product selection beyond query-product relevance signals?}
    \RQ{How do verbosity and relevance interact when they provide conflicting signals to an LLM-based RS?}
\end{itemize}

% We solve these questions by proposing a methodology and apply it to verbosity as use case. Main contributions:
To address these questions, we propose a statistically grounded, label-free methodology for detecting and characterizing structural biases in LLM-based RS, and instantiate it on verbosity bias as a concrete use case. The methodology uses paraphrasing models as a controlled rewriting mechanism: by applying semantics-preserving paraphrasers to candidate descriptions and randomizing both their assignment and their prompt position, we construct a setting in which any systematic deviation from uniform paraphraser selection, after controlling for semantic deviations, constitutes evidence of structural bias. A four-step statistical pipeline then detects the presence of bias, quantifies the independent contribution of each surface-level feature after controlling for relevance, determines the direction of each feature's effect, and characterizes how surface-level and relevance signals interact when they conflict. Our main contributions are the following:

\begin{itemize}
    \item{\textbf{Design-based guarantee for label-free bias detection.} We show that, when paraphrasers are assigned to products at random, a recommender that is invariant to paraphrasing must select every paraphraser with equal probability. This randomization guarantee yields a valid test for structural bias that requires no relevance annotations.}
    \item{\textbf{Extensible bias evaluation framework.} We propose a 
    label-free evaluation framework for structural bias in LLM-based RS that decomposes selection behavior into interpretable surface-level features. New feature extractors can be integrated with a single configuration change.}
    \item{\textbf{Empirical characterization of verbosity bias.} We detect and quantify verbosity bias across 9 recommender LLMs, showing that description length has a positive effect on product selection in all of them after controlling for the measured query-product relevance signals, with a magnitude that varies across model families.}
    \item{\textbf{A practical auditing protocol.} We package the methodology as a step-by-step audit for deployed LLM-based recommenders that requires only black-box access, reuses paraphrases across audits, and comes with sample-size guidance, so that practitioners can monitor structural biases whenever the model, prompt or catalog changes.}
\end{itemize}

% Results description
We apply our methodology to a dataset of $20{,}000$ sampled user queries over a subset of the Amazon ML Challenge 2023 catalog, evaluating nine recommender LLMs spanning five model families. Seven of the nine models select paraphrased descriptions non-uniformly, showing that semantically equivalent rewritings of the same products are not treated as interchangeable. In all nine models, description length has a positive effect on selection after controlling for the measured relevance and semantic similarity signals. Relevance remains the primary driver of the recommenders' decisions: when verbosity and relevance point in opposite directions, the more relevant description is still preferred, but the preference gap is on average less than half that observed when both signals agree, indicating that verbosity acts as a secondary signal that partially counteracts relevance. The strength of this effect varies across model families: the GPT-OSS models are the least sensitive to description length, whereas the DeepSeek-R1 distillations are the most sensitive.

% Paper outline
The remainder of this paper is organized as follows. Section~\ref{sec:relatedwork} reviews related work on structural biases in LLM-based RS, covering both positional and verbosity bias and identifying the gap that motivates our methodology. Section~\ref{sec:framework} introduces the formal evaluation framework and the four-step statistical methodology for label-free structural bias detection. Section~\ref{sec:experiments} instantiates the methodology on verbosity bias, describing the experimental setup and reporting the results across nine recommender LLMs. Section~\ref{sec:conclusions} summarizes the main findings, discusses their implications, and outlines directions for future work.

\section{Related work}
\label{sec:relatedwork}

% RS fundamentals → LLM-based RS architectures
Recommender systems (RS) are designed to surface relevant items from a large catalog by modeling user preferences, typically through collaborative filtering, content-based methods, or hybrid approaches~\cite{10.1145/3285029, BOBADILLA2013109}. At inference time, RS rank candidates using one of three paradigms: pointwise approaches score each item independently, pairwise approaches compare items in isolation, and listwise approaches rank the full candidate set jointly. Recent work has shifted toward LLM-based RS~\cite{lin2025how}, which leverage LLMs as the recommendation core. Unlike traditional methods, LLM-based RS can process rich natural language descriptions of both users and items, enabling zero-shot generalization and flexible integration into conversational pipelines~\cite{li-etal-2024-large}, and they are increasingly combined with retrieval augmentation and collaborative signals~\cite{KIM2026132508, LUO2026129508, hou2024llmrank}. In this setting, LLMs most commonly operate in a listwise fashion, receiving the full candidate list as a prompt and producing a ranked output directly, a paradigm that, as we discuss below, is particularly exposed to structural biases arising from how items are presented in the input.

% Bias taxonomy: fairness/social vs. structural (positioning our scope)
Biases in LLM-based RS can be broadly grouped into two categories. Fairness and social biases arise when model outputs systematically 
disadvantage users or items along protected attributes such as gender, age, or ethnicity~\cite{gallegos_bias_2024}. In LLM-based RS, such biases can also enter through LLM-generated text: Zhang et al.~\cite{10.1145/3736404} show that user and item profiles generated by LLMs to support recommendation inherit the model's propensity biases, which they mitigate through counterfactual inference. Structural biases, by contrast, stem from how LLMs process input at inference time, including sensitivity to presentation order, text length, or surface wording, independently of demographic factors~\cite{10.1145/3564284}. We acknowledge that these categories are not perfectly separable: positional bias, for instance, may partly reflect human tendencies learned during pretraining~\cite{10.1007/978-3-319-78105-1_37}. This paper focuses exclusively on structural biases, which remain undercharacterized in LLM-based RS despite their direct implications for recommendation quality.

% LLM structural biases
A growing body of work has documented that LLMs exhibit systematic behavioral biases when used as evaluators or rankers, independent of the semantic content of the inputs. Recent surveys and bias-quantification frameworks identify a wide range of such limitations, including positional, verbosity, authority, sentiment, and distraction biases, among others~\cite{GU2026101253}. These biases can be understood as sharing a common mechanism: LLMs appear to use surface-level textual features as proxies for quality, a shortcut that is particularly consequential when the model acts as a recommender selecting among retrieved products described in natural language. In this paper we mainly focus on two of these biases, namely positional and verbosity biases.

% Positional bias
Positional bias refers to the tendency of LLMs to favour candidates based on where they appear in the input prompt rather than on their intrinsic relevance. Ground-truth-dependent studies have documented it in question answering, where Liu et al.~\cite{liu-etal-2024-lost} revealed primacy and recency effects on datasets with known answers, and in recommendation, where Hou et al.~\cite{hou2024llmrank} showed that LLM-based rankers evaluated on standard benchmarks are sensitive to candidate order. Label-free approaches avoid relevance annotations by requiring the output to be invariant to input order, either by aggregating outputs over shuffled inputs into an order-independent consensus, as in permutation self-consistency~\cite{tang-etal-2024-found}, or by enforcing invariance inside the model, as Li et al.~\cite{LI2025128726} do by averaging attention weights across permutations. Both treat any change in the output as an error, which presumes a single correct answer. Recommendation lacks such a ground truth: Peláez-González et al.~\cite{pos_bias} model it as a choice within a set of equally relevant items, so switching between them is not by itself evidence of bias, and propose statistically grounded methodologies to detect positional bias in this setting. Positional bias in LLM-based RS thus already has a dedicated detection framework, but other structural biases do not, and detecting them requires more than measuring output sensitivity to input perturbations, since recommendations lack a single ground-truth answer against which such changes could be judged.

% Verbosity bias
Verbosity bias refers to the tendency of LLMs to assign higher scores to longer or more elaborate inputs independently of their actual quality. On the ground-truth side, Dubois et al.~\cite{dubois2024length} showed that the AlpacaEval benchmark systematically favours longer outputs and proposed a length-controlled variant anchored to human preference labels, using regression to estimate the counterfactual win rate had both responses been the same length. Label-free probes for verbosity bias rely on controlled inflation of inputs rather than human annotations. Zheng et al.~\cite{zheng2023judging} proposed a \emph{repetitive-list attack}, which inflates answers by prepending a rephrased version of the original list without adding new information, and observing whether an LLM judge assigns a higher score to the inflated version. Ye et al.~\cite{ye2025justice} likewise include verbosity bias in their consistency-based taxonomy, confirming that length-inflated answers systematically receive higher scores independently of content quality. Closest to our setting, Tang et al.~\cite{tang2026biasrecbench} introduce BiasRecBench, which evaluates LLMs that select one option from a small candidate list in paper review, e-commerce and recruitment, and includes verbosity among the injected biases. Their evaluation relies on constructed ground truth: candidate lists are synthesized with a calibrated quality gap between an optimal option and sub-optimal ones, and bias is measured as the loss of accuracy when a sub-optimal option is rewritten to be more verbose. Despite this evidence, verbosity bias has received considerably less attention than positional bias in the context of LLM-based RS, where item descriptions naturally vary in length, and to the best of our knowledge it has only been evaluated there against constructed ground-truth labels.

% Gap: synthesis of what's missing and how we fill the gap
Despite the growing evidence that both positional and verbosity biases affect LLM outputs, their rigorous characterization in LLM-based RS remains underexplored. A central difficulty is that variability in the recommendation is not by itself evidence of bias: it may reflect legitimate differences in relevance, and equally relevant items may be selected interchangeably without any bias being involved. This distinguishes recommendation from question answering, where a single answer is correct and invariance can be required for each individual input~\cite{LI2025128726}. Existing methodologies resolve this ambiguity through external references: LLM-as-judge settings assess free-form responses against a reference answer, and evaluation frameworks for LLM-based RS such as CFaiRLLM~\cite{10.1145/3725853} compare recommendations against ground-truth user preferences. Such references are not necessarily available in recommendation settings, where items are distinguished only by their natural language descriptions. While positional bias detection has already received a statistically grounded treatment in this setting~\cite{pos_bias}, no equivalent methodology exists for verbosity bias. Concretely, the remaining challenge is to detect and characterize verbosity-related selection patterns in existing recommendation datasets without ground-truth relevance labels, and to assess how these patterns vary across recommender models after accounting for measured relevance signals. This paper addresses these open questions by proposing statistically grounded, label-free methodologies for detecting and quantifying verbosity bias in LLM-based RS.

\section{Methodology for structural bias detection in LLM-based RS}
\label{sec:framework}
% Presents your contribution as a unified whole:
% the controlled testbed + the statistical pipeline together.
% Reviewers understand immediately that both are novel contributions.

\begin{figure}
    \includegraphics[width=\linewidth]{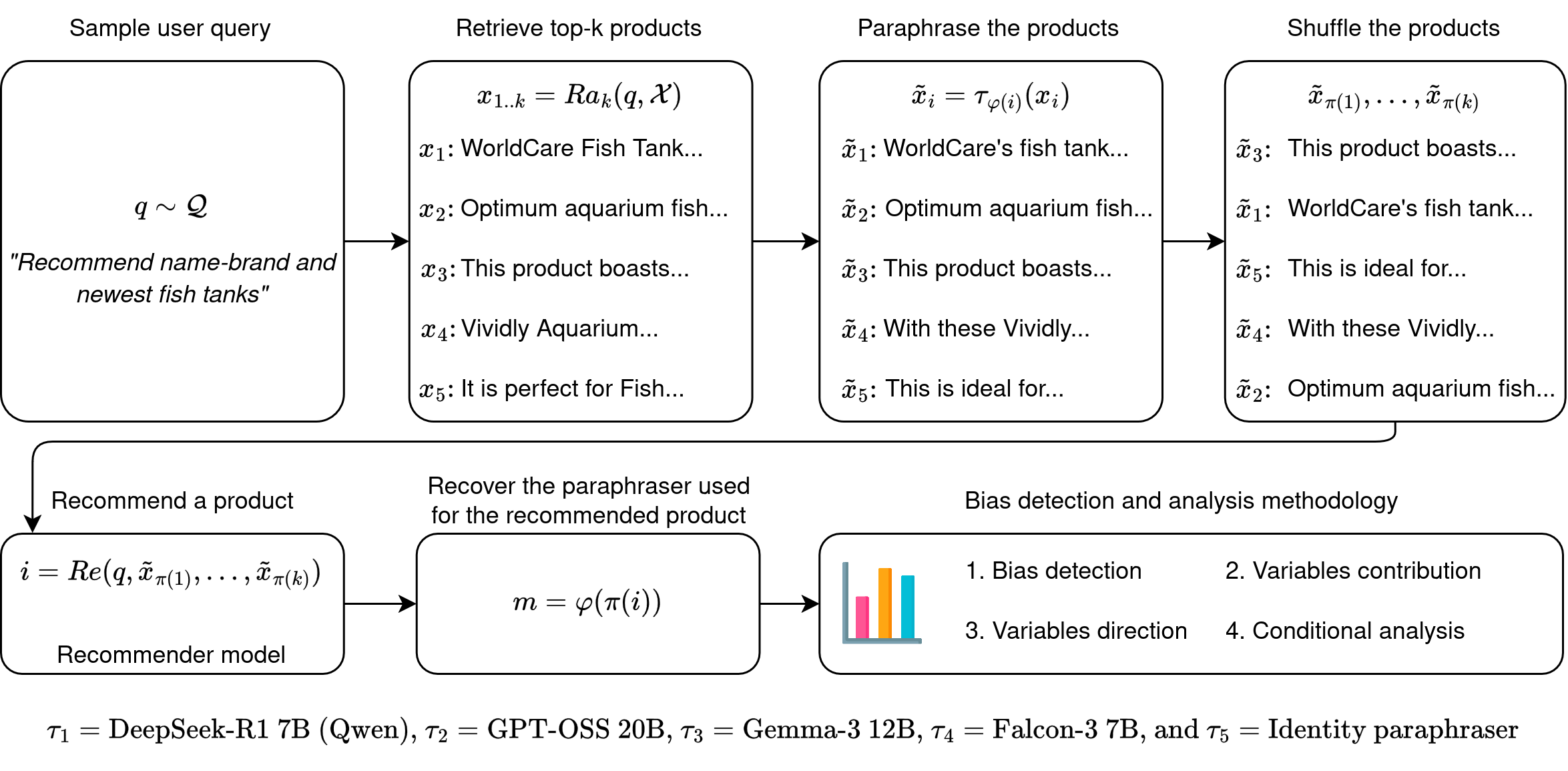}

    \caption{Evaluation framework for structural bias detection. For a sampled query $q$, the retriever $Ra_k$ returns the top-$k$ products $x_1, \ldots, x_k$ from the catalog $\mathcal{X}$. Each product $x_i$ is rewritten by its assigned paraphraser, $\tilde{x}_i = \tau_{\varphi(i)}(x_i)$, and the rewritten products are presented to the recommender in the order $\tilde{x}_{\pi(1)}, \ldots, \tilde{x}_{\pi(k)}$, where $\varphi$ and $\pi$ are independent, uniformly sampled permutations. The recommender returns a prompt position $i$, from which the selected paraphraser is recovered as $m = \varphi(\pi(i))$. In the example shown, $k=5$, the paraphrasers are $\tau_1 = $ DeepSeek-R1 7B (Qwen), $\tau_2 = $ GPT-OSS 20B, $\tau_3 = $ Gemma-3 12B, $\tau_4 = $ Falcon-3 7B and $\tau_5 = $ the identity paraphraser, and the sampled permutations are given by $(\varphi(1), \ldots, \varphi(5)) = (1, 5, 3, 2, 4)$ and $(\pi(1), \ldots, \pi(5)) = (3, 1, 5, 4, 2)$. For instance, if the recommender selects prompt position $i = 2$, the recommended product is $x_{\pi(2)} = x_1$, which was rewritten by $\tau_{\varphi(1)} = \tau_1$, so $m = 1$.}
    \label{fig:evaluation_framework}
\end{figure}

% Introductory paragraph for the full section
This section presents the methodology for detecting and characterizing structural bias in LLM-based RS, summarized in \autoref{fig:evaluation_framework}. First, Section~\ref{sec:framework:testbed} introduces the formal evaluation framework, which extends our prior formalization of LLM-based RS to account for structural biases arising from surface-level features of product descriptions, and defines the probability distribution that enables their systematic detection. Second, Section~\ref{sec:framework:methodology} builds on this framework to introduce a four-stage statistical methodology for detecting, quantifying, and characterizing such biases in a label-free manner.

% Paragraph explaining the evaluation framework
The evaluation framework, introduced in Section~\ref{sec:framework:testbed}, formalizes the notion of surface-level features and establishes the theoretical foundation for their detection. We introduce paraphrasing models as a controlled rewriting mechanism that induces systematic variability in surface-level features while preserving semantic content. We then define a probability distribution over paraphraser selections that, under the null hypothesis of no structural bias and semantic preservation, is provably uniform. Any statistically significant departure from uniformity therefore constitutes evidence that the recommender conditions its decisions on surface-level features of the product descriptions, or that the paraphrasers alter their semantic content (see Remark~\ref{rem:semantic}).

% Paragraph explaining the evaluation methodology
The statistical methodology, introduced in Section~\ref{sec:framework:methodology}, uses the theoretical framework and defines a four-stage pipeline for bias detection and analysis. Starting from the probability distribution defined in the framework, it sequentially tests for the presence of bias via goodness-of-fit testing, quantifies the contribution of surface-level and semantic features via conditional logit modeling, determines the direction of each feature's effect via coefficient estimation, and analyzes the interaction between semantic and surface-level signals via a pairwise conditional decomposition of selection probabilities. Together, these stages provide a complete characterization of structural biases that is both statistically grounded and interpretable.

\subsection{Formal evaluation framework}
\label{sec:framework:testbed}
% Pipeline architecture: query → retriever → paraphraser → recommender

% Introduction to section
This section introduces the formal framework for detecting and evaluating structural biases in LLM-based RS without requiring annotated relevance labels. We begin by reviewing the LLM-based RS formalization and the treatment of positional bias, which serves as the foundation for the present framework. We then extend this formalization to structural biases arising from surface-level features of product descriptions, and define a probability distribution that enables their systematic detection and quantification.

% Introduction to the RS definition and implementation (defined in previous work)
An LLM-based RS implemented as a RAG system consists of two core components: a retriever module $Ra_k$ and a recommender module $Re$. Given a user query $q$ and a product set $\mathcal{X}$, the retriever computes the similarity scores between $q$ and the titles of products in $\mathcal{X}$, returning the top-$k$ ranked products $x = (x_1, \ldots, x_k)$. A qualitative example of retrieved products is shown in \autoref{fig:user_queries_examples}. The recommender then receives $q$ alongside the titles and descriptions of the products, selecting the position of the recommended product as presented in the prompt. 

\begin{figure}
    \begin{tcolorbox}[colback=gray!5!white, colframe=gray, sharp corners, title={Product descriptions retrieved with query "$q = \text{Recommend name-brand and newest fish tanks}$"}]
    \centering
    \begin{itemize}
        \item[$x_1 =$] WorldCare Fish Tank Underwater Aquascaping Artificial Grass Plant Decor Tricolor
        \item[$x_2 =$] Optimum aquarium fish food is suitable for all aquarium fish species such as goldfish,fancy carp,sorddtails,platies,mollies,angelfish,cichlids.
        \item[$x_3 =$] This product boasts advanced auto glass manufacturing for a high-definition display and is designed for smart, effortless operation and cleaning. Its elegant double-arc shape enhances visual appeal, while integrated LED lighting promotes energy efficiency, environmental responsibility, and longevity. Ideal for adding a decorative touch to offices or home desktops, it features a 42x25x46cm tank size with a 32L water capacity. The design is unique, lightweight, user-friendly, and built to last, with minimal energy usage and a beautiful aesthetic.
        \item[$x_4 =$] Creating a beautiful aquarium is easy with these Vividly Aquarium Decor. The Float Plants Aquarium Decor allow the aquarium to become a living place of art, while providing excellent hiding place for fish. They are anchored with a weighted ceramic base, so they will stay where they placed in the tank
        \item[$x_5 =$] It is perfect for Fish Tank Aquarium Live Plant. It comes in a size of 27 cm.
    \end{itemize}
\end{tcolorbox}
    \caption{Products retrieved using the \texttt{BAAI/bge-small-en-v1.5} ranker, where $x_{1..k} = Ra_k(q, \mathcal{X})$ and $\mathcal{X}$ is a subset of the Amazon ML Challenge 2023 dataset. We only display the product descriptions although the products are composed by both title and description.}
    \label{fig:user_queries_examples}
\end{figure}

% Definition of positional bias and how to take it into consideration
A key confound in measuring recommendation bias is positional bias, which is the tendency of $Re$ to favor products at specific prompt positions regardless of their content. To control for this, in \cite{pos_bias} the following experiment was introduced. Let us sample a query $q \sim \mathcal{Q}$ and let $x = (x_1, \ldots, x_k)$ be its top-$k$ ranked products. A permutation $\pi$ is sampled independently and uniformly from $\mathcal{S}_k$, the set of all permutations of $\{1, \ldots, k\}$. 
The permutation $\pi$ reorders the products before they are presented to $Re$, so that the recommender receives $(x_{\pi(1)}, \ldots, x_{\pi(k)})$ rather than $(x_1, \ldots, x_k)$. If the recommender selects prompt position $j$, the recommended product is $x_{\pi(j)}$, so its original index is recovered as $\pi(j)$. The probability that the recommender picks the product at position $i$ in the prompt is then $p(i) = \operatorname{Pr}_{q \sim \mathcal{Q}, \pi \sim U(\mathcal{S}_k)}\left[Re(q, x_{\pi(1)}, \ldots, x_{\pi(k)}) = i\right]$, where the probability is taken over the joint distribution of queries $q$ and permutations $\pi$. The key contribution of \cite{pos_bias} is proving that, if the RS does not suffer from positional bias, then $p(i)$ is the uniform distribution (or close to it under stochastic decoding).

% Introduction to surface-level features of products
Beyond positional bias, recommendation outcomes may be systematically influenced by intrinsic properties of the product descriptions that are independent of their semantic content. We refer to these as  \emph{surface-level features}: measurable textual properties of a product that can be extracted directly from its description without reference to the query or the retrieval context. Formally, let $\phi_j: \mathcal{X} \rightarrow \mathcal{F}_j$ be a feature extraction function mapping a product $x \in \mathcal{X}$ to a feature vector in a mixed feature space $\mathcal{F}_j$, where each $\mathcal{F}_j$ may be continuous (e.g.\ $\mathbb{N}^{+}$ for verbosity, quantified as word or token count) or categorical (e.g.\ syntactic voice).

% Introduction to paraphrasers and their bias in feature space
Detecting structural bias in natural language inputs is inherently challenging: unlike categorical variables, which admit direct group comparisons, or numerical variables, which can be discretized into bins, text descriptions lack natural partitions over which to compare probability distributions. To address this, we propose the use of paraphrasing models as a controlled rewriting mechanism. Formally, a paraphraser is a function $\tau_m: \mathcal{X} \rightarrow \mathcal{X}$ that maps a product description to a semantically equivalent description with potentially different surface-level features. In practice, each paraphraser comes from a distinct LLM that has been asked to paraphrase the product description. We also use the identity paraphraser, which leaves the product description unaltered, in order to determine if the original descriptions are less favored compared to the paraphrased ones.

\begin{figure}
    \begin{tcolorbox}[colback=gray!5!white, colframe=gray, sharp corners, title={Paraphrased product descriptions retrieved with query "$q = $ Recommend name-brand and newest fish tanks"}]
    \centering
    \begin{itemize}[leftmargin=4cm]
        \item[$\tilde{x}_1 = \tau_{\varphi(1)}(x_1) = \tau_{1}(x_1) = $] WorldCare's fish tank features underwater aquascaping with artificial grass decor in a trichromatic design.
        \item[$\tilde{x}_2 = \tau_{\varphi(2)}(x_2) = \tau_{5}(x_2) = $] Optimum aquarium fish food is suitable for all aquarium fish species such as goldfish,fancy carp,sorddtails,platies,mollies,angelfish,cichlids.
        \item[$\tilde{x}_3 = \tau_{\varphi(3)}(x_3) = \tau_{3}(x_3) = $] This product boasts advanced auto glass manufacturing for a high-definition display and is designed for smart, effortless operation and cleaning. Its elegant double-arc shape enhances visual appeal, while integrated LED lighting promotes energy efficiency, environmental responsibility, and longevity. Ideal for adding a decorative touch to offices or home desktops, it features a 42x25x46cm tank size with a 32L water capacity. The design is unique, lightweight, user-friendly, and built to last, with minimal energy usage and a beautiful aesthetic.
        \item[$\tilde{x}_4 = \tau_{\varphi(4)}(x_4) = \tau_{2}(x_4) = $] With these Vividly Aquarium Decor, setting up a stunning aquarium is simple. The Float Plants Aquarium Decor transform the tank into a living work of art while also providing excellent hiding spots for fish. They are secured by a weighted ceramic base, ensuring they stay in place inside the tank.
        \item[$\tilde{x}_5 = \tau_{\varphi(5)}(x_5) = \tau_{4}(x_5) = $] This is ideal for live plant aquariums like Fish Tank. It measures 27 cm in size.
    \end{itemize}
\end{tcolorbox}
    \caption{Product descriptions retrieved as described in \autoref{fig:user_queries_examples} and paraphrased by $\tau_1 = \text{DeepSeek-R1 7B (Qwen)}$, $\tau_2 = \text{GPT-OSS 20B}$, $\tau_3 = \text{Gemma-3 12B}$, $\tau_4 = \text{Falcon-3 7B}$, and $\tau_5 = \text{Identity paraphraser}$. We shuffle the paraphrasers using the sampled permutation $\varphi$ given by $(\varphi(1), \ldots, \varphi(5)) = (1, 5, 3, 2, 4)$, so that product $x_i$ is rewritten by paraphraser $\tau_{\varphi(i)}$.}
    \label{fig:paraphrase_examples}
\end{figure}

% Build the final equation

Given a tuple of paraphraser models $(\tau_1, \ldots, \tau_k)$, one per retrieved product, each retrieved product is rewritten by its assigned paraphraser before being presented to $Re$. However, a naive assignment, where paraphraser $\tau_m$ is always applied to the product ranked $m$-th by $Ra_k$, would conflate the paraphraser's influence with the retriever's relevance ordering, since $Ra_k$ returns products sorted by similarity score and top-ranked products are systematically more relevant. To decouple these two sources of variation, we independently randomize both the assignment of paraphrasers and the order in which products are presented. Specifically, a permutation $\varphi \sim U(\mathcal{S}_k)$ assigns paraphraser $\tau_{\varphi(i)}$ to product $x_i$, yielding the rewritten product $\tilde{x}_i = \tau_{\varphi(i)}(x_i)$, and an independent permutation $\pi \sim U(\mathcal{S}_k)$ determines the prompt order, so that prompt position $j$ shows the product $\tilde{x}_{\pi(j)}$, whose original index is $\pi(j)$. This process is illustrated in \autoref{fig:evaluation_framework}, and a qualitative example of paraphrased products is shown in \autoref{fig:paraphrase_examples}. Since $Re$ returns a prompt position $i$, the recommended product is $x_{\pi(i)}$ and its paraphraser is $\tau_{\varphi(\pi(i))}$. We therefore define the probability of the recommender selecting a product paraphrased by model $m$ as
\begin{equation}
    p(m) =
    \operatorname{Pr}_{
        q \sim \mathcal{Q},\, 
        \varphi \sim U(\mathcal{S}_k),\,
        \pi \sim U(\mathcal{S}_k)
    }\left[
        \varphi\!\left(\pi\!\left(
        Re\!\left(q,\, \tilde{x}_{\pi(1)}, \ldots, 
        \tilde{x}_{\pi(k)}\right)\right)\right) = m
    \right].
    \label{eq:probability_distribution}
\end{equation}

Randomizing $\pi$ makes the prompt position of each product independent of its paraphraser, so that positional bias cannot favor any paraphraser systematically, and randomizing $\varphi$ makes the paraphraser independent of the product's retrieval rank, so that differences in relevance across products cannot favor any paraphraser systematically.

For easier comparison against the uniform baseline, we define the relative selection probability as
\begin{equation}
    p_{\mathrm{rel}}(m) = \left(p(m) - \frac{1}{k}\right) \cdot 100,
    \label{eq:relative_probability_distribution}
\end{equation}
which expresses the deviation of each paraphraser's selection rate from the uniform distribution in percentage points, with positive values indicating above-baseline preference and negative values indicating below-baseline preference.

\begin{figure}
    \includegraphics[width=\linewidth]{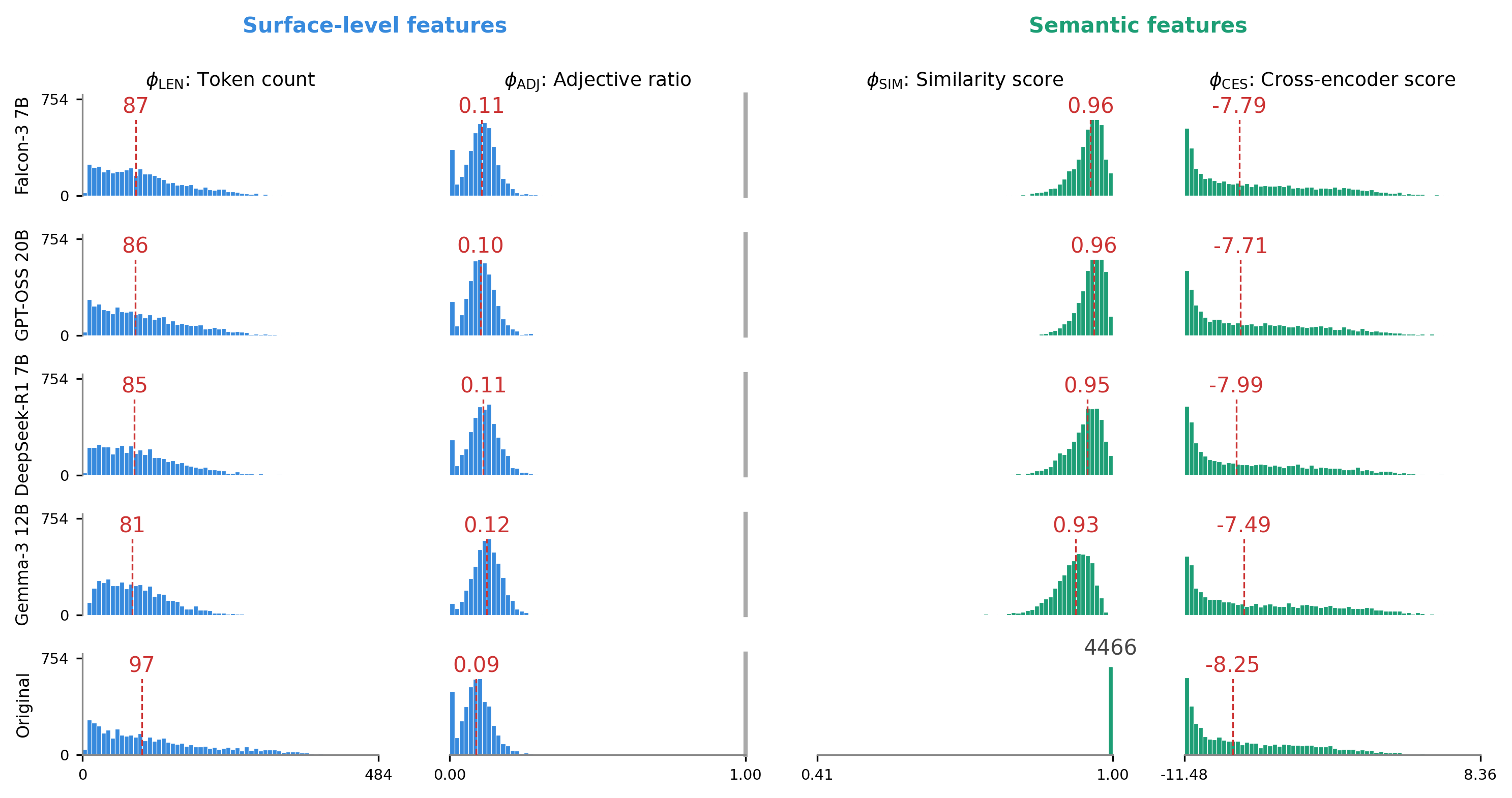}
    \caption{Feature distributions across paraphraser models for the sampled \textit{Amazon ML Challenge 2023} dataset (see \autoref{sec:experiments:setup}). Each column corresponds to a surface-level or semantic feature; each row to a paraphraser. $\phi_{\mathrm{LEN}}$ measures description length in tokens. $\phi_{\mathrm{ADJ}}$ (adjective ratio) is the fraction of words that are adjectives; it is shown as an example of an additional surface-level extractor but is not used in the analysis. $\phi_{\mathrm{CES}}$ measures query--product relevance. $\phi_{\mathrm{SIM}}$ measures paraphrased--original product cosine similarity. The identity paraphraser (\textit{Original}) yields a degenerate $\phi_{\mathrm{SIM}}$ distribution (similarity always 1), shown with a truncated bar. Red dashed lines represent the median.}
    \label{fig:features_distributions}
\end{figure}

We further introduce a theoretical framework for bias decomposition. We consider a features extraction function $\Phi: \mathcal{X} \rightarrow \mathcal{F}_{sem} \times \mathcal{F}_1 \times \ldots \times \mathcal{F}_n$, where $\mathcal{F}_{sem}$ encodes the semantic content of the description and $\mathcal{F}_1 \times \ldots \times \mathcal{F}_n$ encode its surface-level features, such as syntax, wording, or length. Examples of such feature extractors are shown in \autoref{fig:features_distributions}. Concretely, $\Phi(\tau_m(x)) = (\phi_{sem}(\tau_m(x)),\, \phi_{1}(\tau_m(x)),\, \ldots,\, \phi_{n}(\tau_m(x)))$, where $\phi_{sem}$ captures the semantic content of the paraphrased description and $\phi_1, \ldots, \phi_n$ extract its surface-level features. By selecting paraphrasers that induce systematic variability in $\phi_1, \ldots, \phi_n$ while satisfying semantic preservation, it becomes possible to isolate the influence of individual surface-level properties on the recommender's decisions in a label-free manner, and thereby detect structural biases that would otherwise be confounded by semantic content.

% Teaser of the next section. Equation should be uniform
In the following section, we prove that the probability distribution defined in \autoref{eq:probability_distribution} is uniform when the recommender's choice is invariant to paraphrasing, that is, when it does not react to surface-level features and the paraphrasers preserve meaning. Any statistically significant deviation from uniformity therefore signals a violation of at least one of these conditions, which the methodology then disentangles through the analysis of surface-level and semantic features.

\subsection{Uniformity under absence of structural bias}
\label{sec:framework:null}
% Formal claim: under this construction, paraphraser selection 

% Introduction to section
% Define precondition 1 (paraphrasers preserve semantics of the product)
% Define precondition 2 (the recommendation does not depend on the surface-level features)
% Description of hypothesis and proof
%   - Using precondition 2 we can remove surface-level features
%   - Using precondition 1 we can remove the hat
%   - Now the recommendation does not depend on m, and so is a "constant". The shuffle of paraphrasers make the distribution uniform
% Conclusion and considerations (we have to add a semantic preservation signal) (we have to decompose the contribution of relevance and surface-level features)

\begin{intuitionbox}
Suppose the recommender satisfies \emph{No structural bias}, i.e.\ it does not condition its decision on surface-level features such as description length, and \emph{Semantic preservation}, i.e.\ the paraphrasers do not induce a systematic semantic shift. Then the only thing the recommender's decision can depend on is semantic content, which is unaffected by which paraphraser produced it. Since the paraphraser-to-product assignment $\varphi$ is sampled uniformly and independently of this content, the recommender has no way to tell which paraphraser generated the description it selects. With no such signal to exploit, it cannot systematically favor one paraphraser over another, so the identity of the selected paraphraser is governed purely by the uniform randomness of $\varphi$ — and must therefore itself be uniformly distributed. We formalize the structural bias and semantic preservation conditions in Assumption~\ref{ass:structural_bias_and_semantic_preservation}.
\end{intuitionbox}

We defined in the previous section $p(m)$ as the probability of a recommender model selecting a product paraphrased by model $m$, where $m \in \{1, \ldots, k\}$ indexes the set of paraphraser models. This formulation provides a mechanism for detecting structural biases in LLM-based RS. In this section, we prove that $p(m)$ is uniform under Assumption~\ref{ass:structural_bias_and_semantic_preservation}, establishing that any statistically significant deviation from uniformity constitutes evidence that at least one of these conditions fails, that is, either $Re$ conditions its decisions on surface-level features of the product descriptions or paraphrasers alter the recommender's effective decision in a non-uniform way across paraphraser models.

\begin{assumption}[No structural bias and semantic preservation]
\label{ass:structural_bias_and_semantic_preservation}
For every $q$, every retrieved product tuple $(x_1, \ldots, x_k) \in \mathcal{X}^k$, and every paraphraser assignment $\varphi \in \mathcal{S}_k$, we have
\begin{equation*}
    Re(q,\, \tilde{x}_{1}, \ldots, \tilde{x}_{k}) = Re(q,\, x_{1}, \ldots, x_{k}), \qquad \text{ where } \tilde{x}_i = \tau_{\varphi(i)}(x_i),
\end{equation*}
so $\tilde{x}_i$ denotes product $x_i$ rewritten by its assigned paraphraser $\tau_{\varphi(i)}$.

That is, the choice of the recommender system does not depend on the paraphraser used on each product. Intuitively, this occurs when the paraphraser respects the semantic representation of the product description and the structural variation introduced does not affect the choice of the recommender system.
\end{assumption}

\begin{proposition}
\label{prop:uniformity_proposition}
Under Assumption~\ref{ass:structural_bias_and_semantic_preservation}, the probability $p(m)$ defined in \autoref{eq:probability_distribution} is the uniform distribution over $\{1,2,\ldots,k\}$.
\end{proposition}
\begin{proof} 
Fix a query $q$ and a product tuple $(x_1, \ldots, x_k)$, and condition on them throughout. Let $\tilde{x}_i = \tau_{\varphi(i)}(x_i)$ denote product $x_i$ rewritten by its assigned paraphraser. We have
\begin{equation*}
    \begin{split}
    p(m \mid q, x_1, \ldots, x_k) 
    &= \operatorname{Pr}_{\pi,\, \varphi \sim U(\mathcal{S}_k)}\!\left[
        \varphi\!\left(\pi\!\left(
        Re(q,\, \tilde{x}_{\pi(1)}, \ldots, \tilde{x}_{\pi(k)})
        \right)\right) = m
    \right] \\
    &\stackrel{(1)}{=} \operatorname{Pr}_{\pi,\, \varphi \sim U(\mathcal{S}_k)}\!\left[       
        Re(q,\, \tilde{x}_{\pi(1)}, \ldots, \tilde{x}_{\pi(k)}) = \pi^{-1}(\varphi^{-1}(m))
    \right] \\
    &\stackrel{(2)}{=} \operatorname{Pr}_{\pi,\, \varphi \sim U(\mathcal{S}_k)}\!\left[       
        Re(q,\, x_{\pi(1)}, \ldots, x_{\pi(k)}) = \pi^{-1}(\varphi^{-1}(m))
    \right] \\
    &\stackrel{(3)}{=} \sum_{\pi \in \mathcal{S}_k} \frac{1}{k!}\, \operatorname{Pr}_{\varphi \sim U(\mathcal{S}_k)}\!\left[
        Re(q,\, x_{\pi(1)}, \ldots, x_{\pi(k)}) = \pi^{-1}(\varphi^{-1}(m))
    \right] \\        
    &\stackrel{(4)}{=} \sum_{\pi \in \mathcal{S}_k} \frac{1}{k!} \cdot \frac{1}{k}
      = \frac{1}{k},
    \end{split}
\end{equation*}
where 
\begin{enumerate}[label=(\arabic*)]
    \item uses that $\varphi$ and $\pi$ are bijections;
    \item applies Assumption~\ref{ass:structural_bias_and_semantic_preservation} to the reordered tuple $(x_{\pi(1)}, \ldots, x_{\pi(k)})$ with paraphraser assignment $\varphi \circ \pi \in \mathcal{S}_k$, since $\tilde{x}_{\pi(j)} = \tau_{\varphi(\pi(j))}(x_{\pi(j)})$;
    \item conditions on $\pi$, which is uniform over $\mathcal{S}_k$ and independent of $\varphi$;
    \item holds because, for fixed $\pi$, $Re(q,\, x_{\pi(1)}, \ldots, x_{\pi(k)})$ is a fixed number in $\{1, \ldots, k\}$ (or, under stochastic decoding, a random variable independent of $\varphi$), whereas $\varphi^{-1}(m)$ is uniformly distributed over $\{1, \ldots, k\}$ when $\varphi \sim U(\mathcal{S}_k)$, and hence so is $\pi^{-1}(\varphi^{-1}(m))$.
\end{enumerate}
Since this holds for every query and product tuple, the law of total probability gives $p(m) = 1/k$ for all $m \in \{1, 2, \ldots, k\}$.
\end{proof}

\begin{remark}
The proof of Proposition~\ref{prop:uniformity_proposition} holds for every fixed ordering $\pi$: the uniformity of $p(m)$ follows from the random paraphraser assignment $\varphi$ alone. Randomizing $\pi$ is nevertheless essential for the rest of the methodology, since it prevents positional bias from being confounded with the surface-level and semantic features analyzed in Steps~2--4.
\end{remark}

\begin{remark}
The implication from non-uniformity to structural bias admits two non-exclusive diagnostic interpretations, corresponding to violations of Assumption~\ref{ass:structural_bias_and_semantic_preservation}. A violation of Assumption~\ref{ass:structural_bias_and_semantic_preservation} indicates that either $Re$ conditions its decisions on surface-level features of the product descriptions, the central phenomenon studied in this work, or that one or more paraphrasers introduce semantic drift large enough to cross a decision boundary of the recommender. The methodology developed in Section~\ref{sec:framework:methodology} disentangles these two sources by separately quantifying the contributions of semantic and surface-level features to the recommender's decision rule.
\label{rem:semantic}
\end{remark}

\begin{remark} \label{rem:converse}
The converse of Proposition~\ref{prop:uniformity_proposition} does not hold in general: a recommender model may still exhibit structural bias while producing a uniform $p(m)$, for instance if its biases cancel out symmetrically across paraphraser models. Uniformity of $p(m)$ is therefore a necessary but not sufficient condition for the absence of structural bias.
\end{remark}

\begin{remark}\label{rem:general_assignment}
The design couples the number of paraphrasers with the number $k$ of retrieved products only for convenience. If, instead of a permutation $\varphi$, each product $x_i$ is assigned a paraphraser index $a_i$ drawn independently and uniformly from $\{1, \ldots, M\}$, and Assumption~\ref{ass:structural_bias_and_semantic_preservation} holds for every assignment $(a_1, \ldots, a_k) \in \{1, \ldots, M\}^k$, the same argument shows that the selected paraphraser is uniformly distributed over $\{1, \ldots, M\}$. The framework therefore accommodates any number of paraphrasers and candidates.
\end{remark}

\subsection{Statistical structural bias detection methodology}
\label{sec:framework:methodology}

% Introductory paragraph
This section introduces a four-step methodology for detecting and characterizing structural biases in LLM-based RS. The first step tests for the presence of bias by assessing whether the empirical distribution of paraphraser selections deviates significantly from uniformity. The second step quantifies the contribution of each surface-level and semantic feature to the recommender's decision rule, measuring how much each feature contributes to the fit of a choice model. The third step determines the direction of each feature's effect, establishing whether higher values of a given feature increase or decrease the probability of selection. The fourth step analyzes the interaction between the semantic feature and the surface-level feature under study, decomposing selection probabilities by their joint conditions to reveal how the two signals combine or conflict in driving the recommender's decisions.

% Experiment of table 1
\paragraph{Step 1. Bias detection.} Having established in Proposition~\ref{prop:uniformity_proposition} that an unbiased recommender must select paraphrasers uniformly, we now test whether observed selections actually follow this distribution. A significant departure is our first signal that something beyond semantics is driving choices. Under the null hypothesis of no structural bias, the empirical selection counts $\{n_1, \ldots, n_k\}$, where $n_m$ denotes the number of times the recommender selected a product paraphrased by model $m$, follow a multinomial distribution with equal cell probabilities $1/k$. We test this using a Pearson chi-squared goodness-of-fit statistic with $k-1$ degrees of freedom at significance level $\alpha = 0.05$. A rejection, however, does not by itself identify the source of non-uniformity: it may reflect genuine sensitivity of $Re$ to surface-level features, semantic drift introduced by one or more paraphrasers, or both. The following steps disentangle these contributions.

% Experiment of table 2
\paragraph{Step 2. Variables contribution.} Knowing that bias exists is not enough: a significant $\chi^2$ result could reflect sensitivity to description length, to subtle semantic drift introduced by paraphrasers, or to both simultaneously. This step asks how much each feature independently contributes to explaining the recommender’s choices, separating the contribution of surface-level features from that of semantic ones. To enable the application of a conditional logit model, we construct a long-format dataset $\mathcal{D}$ of size $k \times N$, where $N$ is the number of queries and $k$ is a shared count of candidate products and paraphraser models. Each query $q$ contributes exactly $k$ rows to $\mathcal{D}$, one per candidate paraphraser, forming a query group. Within each group, each row corresponds to a specific paraphraser $m \in \{1, \ldots, k\}$ and contains the surface-level feature vector $(\phi_1(x_{q,m}), \ldots, \phi_n(x_{q,m}))$ and the semantic feature vector $(\phi_{sem,1}(x_{q,m}), \ldots, \phi_{sem,s}(x_{q,m}))$ of the corresponding paraphrased description, alongside a binary indicator $y_{q,m} \in \{0, 1\}$ encoding whether paraphraser $m$ was selected for query $q$. By construction, exactly one row per group satisfies $y_{q,m} = 1$, forming a one-hot encoding of the recommender's choice within each query group. This structure is directly compatible with the softmax likelihood of the conditional logit model. Prior to model fitting, all features are standardized globally across all $k \times N$ rows, which preserves the relative ordering of features across products within each query while placing all variables on a common scale.

To quantify the contribution of each feature to the recommender's decision, we fit a conditional logit model~\cite{mcfadden} grouped by query $q$. The model assumes that the probability of the recommender selecting paraphraser $m$ for query $q$ follows a softmax over a linear combination of the feature vectors:
\begin{equation*}
    \Pr[y_{q,m} = 1 \mid q] = 
    \frac{e^{x_{q,m}^{\top} \beta}}{\sum_{j=1}^{k} e^{x_{q,j}^{\top} \beta}},
\end{equation*}
where $x_{q,m}$ is the feature vector of paraphraser $m$ for query $q$ and $\beta$ is the vector of coefficients to be estimated. The model is fit by maximizing the conditional log-likelihood over all $N$ query groups:
\begin{equation*}
    \ell(\beta) = \sum_{q=1}^{N} \left[
        x_{q, m^*_q}^{\top} \beta - 
        \ln \sum_{j=1}^{k} e^{x_{q,j}^{\top} \beta}
    \right],
\end{equation*}
where $m^*_q = \arg\max_m y_{q,m}$ denotes the paraphraser selected by the recommender for query $q$. The resulting estimator $\hat{\beta} = \arg\max_{\beta}\, \ell(\beta)$ is obtained via standard maximum likelihood estimation.

To assess the contribution of each feature to the model's predictive power, we adopt a McFadden $R^2$ measure~\cite{mcfadden}, defined as
\begin{equation*}
    R^2 = 1 - \frac{\ell(\hat{\beta})}{-N \cdot \ln k},
    \label{eq:r2}
\end{equation*}
where $-N \cdot \ln k$ is the log-likelihood of the null model under a uniform distribution over $k$ paraphrasers, serving as a maximum-entropy baseline. Note that McFadden's $R^2$ is a likelihood-based measure of fit rather than a proportion of explained variance, and its values are typically much lower than those of the linear-regression $R^2$. We consider two decompositions of $R^2$. The solo contribution of feature $i$ is obtained by fitting the conditional logit using only $\phi_i$ as a predictor:
\begin{equation}
    R^2_{\text{solo},i} = 1 - \frac{\ell(\hat{\beta}_i)}{-N \cdot \ln k}, 
    \quad x \in \mathcal{F}_i.
    \label{eq:r2_solo}
\end{equation}
The unique contribution of feature $i$ measures the additional explanatory power attributable exclusively to $\phi_i$, after accounting for all other features:
\begin{equation}
    R^2_{\text{unique},i} = R^2_{\text{joint}} - R^2_{\text{excl},i},
    \label{eq:r2_unique}
\end{equation}
where $R^2_{\text{joint}}$ is the $R^2$ of the full model fitted on $\mathcal{F}_{sem} \times \mathcal{F}_1 \times \cdots \times \mathcal{F}_n$, and $R^2_{\text{excl},i}$ is the $R^2$ of the model fitted on all features except $\phi_i$, i.e.\ on $\mathcal{F}_{sem} \times \mathcal{F}_1 \times \cdots \times \mathcal{F}_{i-1} \times \mathcal{F}_{i+1} \times \cdots \times \mathcal{F}_n$.

\paragraph{Step 3. Variables direction.} $R^2$ tells us that description length helps explain selection, but not whether longer descriptions are favored or penalized. This step recovers the sign of each feature's effect, turning the $R^2$ decomposition of the previous step into a directional claim about the recommender's behavior. For this purpose, we examine the estimated coefficients $\hat{\beta}$ of the conditional logit model. A positive coefficient $\hat{\beta}_i > 0$ indicates that higher values of feature $\phi_i$ increase the probability of the corresponding paraphraser being selected by the recommender, while a negative coefficient $\hat{\beta}_i < 0$ indicates the opposite. The full coefficient vector is obtained as:
\begin{equation*}
    \hat{\beta} = \underset{\beta}{\arg\max}\sum_{q=1}^{N}\left[
        x_{q, m^*_q}^{\top} \beta - 
        \ln \sum_{j=1}^{k} e^{x_{q,j}^{\top} \beta}
    \right],
    \label{eq:beta}
\end{equation*}
where the feature vector $x_{q,m}$ may include any combination of surface-level and semantic features, depending on the configuration in which the model is fitted. Optimization is performed using the Broyden-Fletcher-Goldfarb-Shanno (BFGS) algorithm.

Analogously to the $R^2$ decomposition, we consider two configurations for estimating the coefficient of feature $\phi_i$. The solo coefficient $\hat{\beta}^{\text{solo}}_i$ is obtained by fitting the conditional logit using only $\phi_i$ as a predictor:
\begin{equation}
    \hat{\beta}^{\text{solo}}_i = \hat{\beta}_1, 
    \quad x \in \mathcal{F}_i,
    \label{eq:beta_solo}
\end{equation}
where $\hat{\beta}_1$ denotes the single coefficient estimated in this univariate model. The unique coefficient $\hat{\beta}^{\text{unique}}_i$ is obtained by fitting the conditional logit on the full feature space $\mathcal{F}_{sem} \times \mathcal{F}_1 \times \cdots \times \mathcal{F}_n$ and retaining the coefficient associated with $\phi_i$:
\begin{equation}
    \hat{\beta}^{\text{unique}}_i = \hat{\beta}_i, 
    \quad x \in \mathcal{F}_{sem} \times \mathcal{F}_1 \times 
    \cdots \times \mathcal{F}_n.
    \label{eq:beta_unique}
\end{equation}
Comparing $\hat{\beta}^{\text{solo}}_i$ and $\hat{\beta}^{\text{unique}}_i$ reveals the extent to which the raw effect of $\phi_i$ is attenuated when controlling for other features, thereby isolating its independent contribution to the recommender's decision rule.

\paragraph{Step 4. Conditional analysis of conflicting signals.} Steps 1--3 establish that both semantic and surface-level features independently influence selection. What remains unknown is how the recommender behaves when these two signals conflict, for instance when one description is longer but less relevant than another. To study this, we take the identity paraphraser as the focal paraphraser $f$, so that one side of every comparison is an original description, and compare it in turn with each of the other paraphrasers $c \neq f$.

As in Step~2, $x_{q,m}$ denotes the product retrieved for query $q$ that is paraphrased by paraphraser $\tau_m$. For each query $q$, we compare $x_{q,f}$, the product associated with $q$ and paraphrased by the identity paraphraser, i.e.\ shown with its original description, with $x_{q,c}$, the product associated with $q$ and paraphrased by $\tau_c$. Since paraphrasers are assigned to products by a permutation, these are two different products retrieved for the same query. For each feature $\phi$, the two descriptions are compared through a difference $\Delta$ on a natural log scale: the log-ratio $\log\left(\phi(x_{q,f}) / \phi(x_{q,c})\right)$ for strictly positive features such as length, and the plain difference $\phi(x_{q,f}) - \phi(x_{q,c})$ for scores that are already on a log scale, such as the logits of a relevance model. A positive $\Delta$ means that the original description has the higher value. Both differences are antisymmetric, as swapping the two descriptions only flips the sign, and comparable across queries regardless of the absolute value of the feature. The pair is labeled $+$ if $\Delta > \log(1.1)$, $-$ if $\Delta < -\log(1.1)$, and $=$ otherwise, that is, when the two values differ by less than a factor of $1.1$.

A joint condition $(\mathcal{C}_{\phi_i}, \mathcal{C}_{\phi_{sem}}) \in \{+, -, =\}^2$ fixes the label of the surface-level feature under study and that of the semantic feature. For each comparison paraphraser $c$, we keep the queries whose pair $(x_{q,f}, x_{q,c})$ satisfies the condition and, on these queries, compute the selection gap
\begin{equation}
    \Delta p_{\mathcal{C}_{\phi_i}, \mathcal{C}_{\phi_{sem}}}(f, c) =
    p(m{=}f \mid \mathcal{C}_{\phi_i}, \mathcal{C}_{\phi_{sem}}) -
    p(m{=}c \mid \mathcal{C}_{\phi_i}, \mathcal{C}_{\phi_{sem}}),
    \label{eq:conditional_gap}
\end{equation}
where the two probabilities are the fractions of the kept queries in which the recommender selected the original description and the description rewritten by $\tau_c$, respectively. They need not sum to one, since the recommender may select any of the other $k-2$ candidates. A positive gap means that, under the given condition, the original description is selected more often than the paraphrased one. The reported value averages this gap over all comparison paraphrasers:
\begin{equation}
    \overline{\Delta p}_{\mathcal{C}_{\phi_i}, \mathcal{C}_{\phi_{sem}}} =
    \frac{1}{k-1}
    \sum_{\substack{c=1 \\ c \neq f}}^{k}
    \Delta p_{\mathcal{C}_{\phi_i}, \mathcal{C}_{\phi_{sem}}}(f, c).
    \label{eq:conditional_gap_avg}
\end{equation}

Conditions where both signals are aligned, i.e.\ $(+,+)$ and $(-,-)$, serve as sanity checks: a positive gap under $(+,+)$ and a negative gap under $(-,-)$ confirm that the recommender responds consistently when both features point in the same direction. More informative are the conflicting conditions $(+,-)$ and $(-,+)$, where the two features point in opposite directions, revealing whether the recommender systematically favors one signal over the other when they conflict. Finally, conditions where one signal is approximately tied, i.e. \((C_{\phi_i}, =)\) and \((=, C_{\phi_{\mathrm{sem}}})\), characterize selection patterns when the candidates have similar values of one measured feature but differ in the other. These comparisons provide descriptive evidence about the association of each feature with selection.

% Conclusive paragraph
Together, these four steps form a robust methodology for the detection and analysis of structural biases arising from measurable surface-level features of product descriptions. By grounding each stage in a principled statistical framework and controlling for semantic content throughout, the methodology enables the attribution of observed biases to specific textual properties of the descriptions, after controlling for the measured semantic and relevance signals. Furthermore, its generality makes it applicable to any surface-level feature that can be extracted from product descriptions, beyond the verbosity bias studied in the present work.

\subsection{Auditing protocol for deployed recommender systems}
\label{sec:framework:protocol}

Algorithm~\ref{alg:audit} summarizes the methodology as an audit that can be run on a deployed LLM-based RS. It treats the recommender as a black box: it only needs to submit a prompt and read the selected product, which is the same interface used in production, so the audit can be run with the deployed model, prompt template and catalog. It requires no relevance annotations, so it can be applied to logged production queries as well as to synthetic ones.

\begin{algorithm}[H]
\caption{Label-free audit of structural bias in an LLM-based RS}
\label{alg:audit}
\KwIn{recommender $Re$ (black box), retriever $Ra_k$, catalog $\mathcal{X}$, queries $q_1, \ldots, q_N$, paraphrasers $\tau_1, \ldots, \tau_k$ (including the identity), feature extractors $\phi_{sem}, \phi_1, \ldots, \phi_n$}
\KwOut{bias test, feature contributions and directions, conditional gaps}
\For{$\ell = 1, \ldots, N$}{
    $(x_1, \ldots, x_k) \leftarrow Ra_k(q_\ell, \mathcal{X})$\;
    sample $\varphi, \pi \sim U(\mathcal{S}_k)$ independently\;
    $\tilde{x}_i \leftarrow \tau_{\varphi(i)}(x_i)$ for $i = 1, \ldots, k$ \tcp*{cached across audits}
    $j \leftarrow Re(q_\ell, \tilde{x}_{\pi(1)}, \ldots, \tilde{x}_{\pi(k)})$\;
    record $m_\ell = \varphi(\pi(j))$ and the features of $\tilde{x}_1, \ldots, \tilde{x}_k$\;
}
\textbf{Step 1:} $\chi^2$ goodness-of-fit test of $(m_1, \ldots, m_N)$ against $U_k$\;
\textbf{Step 2:} fit conditional logits; report solo and unique $R^2$\;
\textbf{Step 3:} report the signs and magnitudes of the unique coefficients\;
\textbf{Step 4:} report conditional selection gaps for the feature of interest versus relevance\;
\end{algorithm}

\paragraph{Cost.} An audit of $N$ queries requires exactly $N$ calls to the recommender and at most $N(k-1)$ calls to the paraphrasing models, since the identity paraphraser is free. Paraphrases depend only on the products and the assignment $\varphi$, not on the recommender, so they can be generated once and reused. In our experiments, a single paraphrased dataset was reused for all nine recommenders, so auditing an additional model, or a new version of the prompt, costs only $N$ recommender calls. Feature extraction relies on small encoder models, whose cost is negligible compared with LLM inference.

\paragraph{Sample size.} The required number of queries depends on the strength of the bias and on the step. The length effect estimated in Step~2 is detectable with few queries: since the likelihood-ratio statistic grows linearly with $N$, the unique contributions reported in \autoref{tab:variables_contribution} imply that roughly $900$ queries would give $80\%$ power at $\alpha = 0.05$ for the least length-sensitive model (GPT-OSS~120B), and about $110$ for the most sensitive one (DeepSeek-R1~7B Qwen). The global uniformity test of Step~1 is more demanding, because it aggregates small deviations spread across paraphrasers. Using the selection probabilities in \autoref{tab:uniformity_test} as effect sizes, about $8{,}000$--$10{,}000$ queries give $80\%$ power for the most biased models, and about $19{,}000$ for Gemma-3~27B. These figures treat the estimated effects as true values and are therefore indicative, but they show that an audit with a few thousand queries suffices to characterize verbosity bias, and $10{,}000$--$20{,}000$ queries suffice to detect global non-uniformity of the magnitude observed here.

\paragraph{Extensibility.} New surface-level features, such as sentiment, adjective density, brand mentions or formatting, are added as additional extractors $\phi_i$ without changing the protocol, and the same paraphrased data can be reanalyzed for them at no additional inference cost.

\section{Verbosity bias detection as a use case: results and analysis}
\label{sec:experiments}

This section instantiates the methodology introduced in \autoref{sec:framework:methodology} on a concrete use case: the detection and characterization of verbosity bias in LLM-based RS.  First, \autoref{sec:experiments:setup} describes the experimental setup, detailing the surface-level and semantic features under study, the dataset and retrieval pipeline, and the nine recommender models evaluated. Then \autoref{sec:experiments:results} applies the four-step methodology to this setup and reports the results, consistently finding that query--product relevance is the dominant driver of product selection while description length operates as a consistent secondary signal across Steps~2--4 and the large majority of evaluated models.

\subsection{Instance of our framework and experimental setup}
\label{sec:experiments:setup}
% Dataset: queries, product catalog, retriever, paraphrasers
% Models: list, hyperparameters, prompt template
% Keep this compact as it instantiates the framework, not a contribution
% Experimental setup
% 1. Dataset description
%     - Queries description
%     - Products description
%     - Used paraphrasers [choose criteria]
%     - Retriever
% 2. Evaluated model
%     - List of models
%     - Hyperparameters (temperature, other details)

% Introduction to section
Instantiating our proposed framework requires three main steps. First, we define the surface-level features under study: as a use case, we evaluate verbosity bias, i.e., the tendency of a recommender model to favour products with longer descriptions independently of their relevance. Second, we introduce semantic features to control for shifts in semantics induced by the paraphrasing models. Third, we describe the evaluation dataset and the recommendation system used to study the recommender module.

\paragraph{Surface-level features.} As a case study, we measure the influence of product description length as a surface-level feature, defining the feature extractor $\phi_{\text{LEN}}(x) \in \mathbb{N}$ as the token count of a product description. Concretely, we tokenize each description using the \texttt{en\_core\_web\_sm} pipeline and count all tokens that are neither spaces nor punctuation marks.

% Semantics feature extractors
\paragraph{Semantic features.} As established in \autoref{sec:framework:null}, the methodology must control for semantic shifts induced by the paraphrasing models. Directly measuring the semantic content of a product description is non-trivial, so we rely on three complementary deep-learning based proxies, described below. They are less interpretable than rule-based extractors, but combining three of them provides a more robust control for semantic shifts.

\begin{itemize}
    \item $\phi_{\text{SIM}}(x)$: the cosine similarity between the embeddings of the original and paraphrased product descriptions, computed using the \texttt{BAAI/bge-small-en-v1.5}~\cite{xiao2024c} model. This score lies in $[-1, 1]$, where $1$ indicates identical representations, and serves as the most direct measure of semantic preservation.
    \item $\phi_{\text{RAN}}(q, x)$: the cosine similarity between the embeddings of the user query and the paraphrased product description, computed with the same model. This score measures query-product relevance and is used to control for shifts in how well the paraphrased description matches the query.
    \item $\phi_{\text{CES}}(q, x)$: a query-product relevance score produced by the \texttt{cross-encoder/ms-marco-MiniLM-L6-v2} model, a version of BERT fine-tuned on the \texttt{MS~MARCO} passage ranking dataset. Unlike $\phi_{\text{RAN}}$, this model was trained specifically for information retrieval and jointly encodes the query and description rather than comparing their embeddings independently.
\end{itemize}

\begin{llmprompt}[\linewidth]{Paraphrasing prompt}{prompt:paraphrasing}
Paraphrase the following text. Do not change the meaning of the original text. Provide your answer in a JSON format. Use the format \textasciigrave\{\{"paraphrased": "your answer"\}\}\textasciigrave\newline

\{text\}
\end{llmprompt}

\paragraph{Evaluation framework.} The empirical evaluation follows directly the framework proposed in \autoref{fig:evaluation_framework}. The dataset construction and evaluation process proceeds in six steps.

\begin{enumerate}
    \item \textbf{Sample user queries,} $q \sim \mathcal{Q}$. We sample $20{,}000$ queries by uniformly drawing a query template, a product category, an associated product name, and zero or more attributes. The query space comprises 54 templates, 18 product categories, 9--18 products per category, 9 attribute categories, and 6--15 attributes per attribute category, yielding a population of $69{,}071{,}332$ distinct queries. Invalid category--attribute combinations (e.g.\ the attribute \textit{reliable} is not valid for the category \textit{jewellery}) were removed by means of a manually curated dictionary mapping each attribute to its valid product categories. We believe the resulting query space is sufficiently diverse to support a representative evaluation.
    \item \textbf{Retrieve top-$k$ products,} $x_{1:k} = Ra_k(q, \mathcal{X})$. For each query, we retrieve the top-$k$ products from the \textit{Amazon ML Challenge 2023} catalog\footnote{\url{https://www.kaggle.com/datasets/ashisparida/amazon-ml-challenge-2023}} using \texttt{BAAI/bge-small-en-v1.5} embeddings and cosine similarity. To reduce computational cost, we first filter out products with descriptions shorter than 40 characters and uniformly sample $50{,}000$ products from the remainder, forming the catalog $\mathcal{X}$. The ranker is then applied to this catalog to retrieve the top-$k$ products for each query. We set $k = 5$, since the design requires one paraphraser per retrieved product: this yields four LLM paraphrasers plus the identity paraphraser, while keeping the cost of evaluating $20{,}000$ queries across nine recommenders manageable.
    \item \textbf{Paraphrase the products,} $\tilde{x}_i = \tau_{\varphi(i)}(x_i),\ \varphi \sim U(\mathcal{S}_k)$. Each retrieved product is rewritten either by one of four LLM paraphrasers, namely DeepSeek-R1~7B (Qwen)~\cite{Guo_2025}, GPT-OSS~20B~\cite{openai2025gptoss120bgptoss20bmodel}, Gemma-3~12B~\cite{gemmateam2025gemma3technicalreport} and Falcon-3~7B~\cite{Falcon3}, or by the identity paraphraser. All paraphrasing models receive the same prompt, shown in \autoref{prompt:paraphrasing}, which asks them to rewrite the description while preserving its meaning. The assignment of paraphrasers to products is drawn independently and uniformly for each query via $\varphi$. All paraphrasing models are run at temperature~0 for deterministic outputs.
    \item \textbf{Shuffle the products,} $\tilde{x}_{\pi(1)}, \ldots, \tilde{x}_{\pi(k)}$. The paraphrased products are reordered by an independent uniform permutation $\pi$ before being presented to the recommender, preventing positional bias from acting as a confound.
    \item \textbf{Recommend a product,} $i = Re(q, \tilde{x}_{\pi(1)}, \ldots, \tilde{x}_{\pi(k)})$. We evaluate 9 recommender models across different families and sizes: DeepSeek-R1~7B (Qwen) and~8B (Llama)~\cite{Guo_2025}, Falcon-3~10B and~7B~\cite{Falcon3}, GPT-OSS~120B and~20B~\cite{openai2025gptoss120bgptoss20bmodel}, Gemma-3~12B and~27B~\cite{gemmateam2025gemma3technicalreport}, and Llama-3.1~8B~\cite{grattafiori2024llama3herdmodels}, each at temperature~0. Deterministic decoding ensures reproducibility; moreover, positional preferences in LLM-based RS have been shown to remain stable across decoding temperatures between $0$ and $0.6$~\cite{pos_bias}.
   \item \textbf{Recover the chosen paraphraser,} $m = \varphi(\pi(i))$. The recommended prompt position $i$ is mapped to its original product index $\pi(i)$, and then to the paraphraser assigned to that product, $\varphi(\pi(i))$. The resulting paraphraser indices across all queries form the dataset used to detect and analyze structural biases.
\end{enumerate}

The prompt template used for recommendation, together with all code required to reproduce the dataset and the analysis, is available in our GitHub repository.
 
% Introduction to next section
Having defined the surface-level and semantic features, constructed the query and product dataset, and collected the recommender model outputs, all inputs required by the methodology of \autoref{sec:framework:methodology} are in place. The following section reports the results of applying this methodology to characterize verbosity bias across the nine evaluated recommender models.

\subsection{Results and analysis on verbosity bias detection}
\label{sec:experiments:results}

% Summary of what this test does and where are the results
% Observations and analysis on the data
% Results on this experiment/step
We now apply the four-step methodology of \autoref{sec:framework:methodology} to characterize verbosity bias across the nine evaluated recommender models. Steps~1 through~3 progressively establish the existence, magnitude, and direction of the bias: Step~1 tests whether paraphraser selection deviates significantly from uniformity, Step~2 quantifies how much of the model’s explanatory power is independently attributable to description length after controlling for relevance and semantic preservation, and Step~3 determines the sign of this independent effect. Step~4 then moves beyond aggregate statistics to examine how verbosity and relevance interact across different relative feature conditions, revealing the conditions under which each signal dominates and how they combine when they conflict.

\subsubsection{Bias detection}

We first apply the bias detection step of \autoref{sec:framework:methodology}. Under the null hypothesis of no structural bias and no semantic shift, the paraphraser selection probabilities defined in \autoref{eq:probability_distribution} should be uniform across all paraphrasers. We assess this via a $\chi^2$ goodness-of-fit test, reporting relative selection probabilities as defined in \autoref{eq:relative_probability_distribution} to facilitate comparison against the uniform baseline. Results are shown in \autoref{tab:uniformity_test}.

Seven out of nine recommender models exhibit a statistically significant departure from uniformity ($p < 0.05$), providing evidence of structural bias or semantic shift in most evaluated models. Across models, Gemma-3 12B is the most consistently preferred paraphraser, while DeepSeek-R1 and Falcon-3 are the most consistently avoided. The two GPT-OSS models do not reach statistical significance, suggesting either an absence of bias or that opposing effects cancel out under this configuration.

The magnitude of the observed deviations varies considerably across recommender models. The largest departures are exhibited by DeepSeek-R1~7B (Qwen) and Falcon-3~7B, with relative probabilities reaching $\pm1.3$ percentage points, while GPT-OSS models show the smallest deviations overall. Notably, the identity paraphraser (Original) does not exhibit a consistent pattern: it is among the most preferred paraphrasers for some models (e.g.\ DeepSeek-R1~7B Qwen, $+1.06\%$) and among the most avoided for others (e.g.\ Falcon-3~7B, $-1.09\%$), suggesting that the bias is model-specific rather than a universal preference for or against unmodified descriptions. Within model families, the sign of the deviations is generally consistent across sizes, though the magnitude tends to differ, as observed for both Gemma-3 and Falcon-3 variants.

Since several paraphrasers belong to the same model families as the evaluated recommenders, a possible confound is self-preference, i.e.\ recommenders favoring text generated by their own family. \autoref{tab:uniformity_test} does not support this explanation: both DeepSeek-R1 recommenders avoid the DeepSeek-R1 paraphraser ($-1.13\%$ and $-0.81\%$), the Falcon-3 recommenders show slightly negative deviations for the Falcon-3 paraphraser, and the GPT-OSS recommenders show near-zero deviations for the GPT-OSS paraphraser. Conversely, the Gemma-3 paraphraser is preferred by almost all recommenders, not only by those of the Gemma family, suggesting that the observed preferences are driven by properties of the paraphrased text rather than by family affinity.

These results indicate that most recommender models do not treat semantically equivalent product descriptions uniformly, which motivates the subsequent steps of the methodology to disentangle the contributions of surface-level and semantic features to this observed non-uniformity.

\begin{table}
    \centering
    \begin{tabular}{l|r|r|r|r|r|r}
        \toprule
        \multirow{2}{*}{\textbf{Recommender LLM}} & \multicolumn{5}{c|}{\textbf{Paraphrasers}} & \multirow{2}{*}{\textbf{$p$-value}} \\
         & \textit{DeepSeek-R1} & \textit{Falcon-3} & \textit{GPT-OSS} & \textit{Gemma-3} & \textit{Original} \\
        \midrule
        DeepSeek-R1 7B (Qwen) & \underline{\cellcolor{negred}-1.13\%} & \cellcolor{negred}-0.53\% & \cellcolor{posgreen}+0.11\% & \cellcolor{posgreen}+0.49\% & \cellcolor{posgreen}+1.06\% & \textbf{0.000029} \\
        DeepSeek-R1 8B (Llama) & \cellcolor{negred}-0.81\% & \cellcolor{negred}-0.62\% & \underline{\cellcolor{posgreen}+1.18\%} & \cellcolor{posgreen}+0.46\% & \cellcolor{negred}-0.21\% & \textbf{0.000025} \\
        Falcon-3 10B & \cellcolor{posgreen}+0.19\% & \cellcolor{negred}-0.20\% & \cellcolor{posgreen}+0.43\% & \cellcolor{posgreen}+0.85\% & \underline{\cellcolor{negred}-1.28\%} & \textbf{0.000050} \\
        Falcon-3 7B & \cellcolor{negred}-0.20\% & \cellcolor{negred}-0.01\% & \cellcolor{posgreen}+0.03\% & \underline{\cellcolor{posgreen}+1.28\%} & \cellcolor{negred}-1.09\% & \textbf{0.000018} \\
        GPT-OSS 120B & \cellcolor{negred}-0.12\% & \cellcolor{negred}-0.07\% & \cellcolor{posgreen}+0.14\% & \underline{\cellcolor{posgreen}+0.69\%} & \cellcolor{negred}-0.64\% & 0.059720 \\
        GPT-OSS 20B & \cellcolor{negred}-0.12\% & \cellcolor{negred}-0.27\% & \cellcolor{negred}-0.00\% & \underline{\cellcolor{posgreen}+0.46\%} & \cellcolor{negred}-0.07\% & 0.562556 \\
        Gemma-3 12B & \cellcolor{negred}-0.60\% & \cellcolor{negred}-0.01\% & \cellcolor{posgreen}+0.61\% & \cellcolor{posgreen}+0.64\% & \underline{\cellcolor{negred}-0.65\%} & \textbf{0.003874} \\
        Gemma-3 27B & \cellcolor{negred}-0.56\% & \cellcolor{negred}-0.18\% & \cellcolor{posgreen}+0.53\% & \underline{\cellcolor{posgreen}+0.66\%} & \cellcolor{negred}-0.45\% & \textbf{0.014408} \\
        Llama-3.1 8B & \cellcolor{negred}-0.29\% & \cellcolor{negred}-0.16\% & \cellcolor{posgreen}+0.26\% & \underline{\cellcolor{posgreen}+1.16\%} & \cellcolor{negred}-0.97\% & \textbf{0.000075} \\
        \bottomrule
    \end{tabular}
    \vspace{5pt}
    \caption{The differences $p(m)-1/k$ in our experiments, where the recommender LLM chooses products paraphrased by $k$ specific models ($k=5$), as specified in \autoref{eq:relative_probability_distribution}. A Chi-squared test is performed on the probability of choosing a paraphraser over the uniform distribution, whose results are the provided $p$-values. $p$-value lower than 0.05 (bold) indicates a bias.}
    \label{tab:uniformity_test}
\end{table}

\subsubsection{Variables contribution}

% Summary on what is this step about and present results (reference table, one sentence)
This step quantifies the contribution of each surface-level and semantic feature to the recommender's decision rule, measuring how much each feature contributes to explaining paraphraser selection independently of the others. \autoref{tab:variables_contribution} reports the solo and unique $R^2$ values for all four features ($\phi_{\text{CES}}$, $\phi_{\text{RAN}}$, $\phi_{\text{SIM}}$, and $\phi_{\text{LEN}}$) across all nine recommender models, revealing that query--product relevance is the dominant driver of selection while description length retains an independent contribution after controlling for all other features.

\begin{table}
    \centering
    \begin{tabular}{l|cc|cc|cc|cc}
        \toprule
        \multirow{2}{*}{\textbf{Recommender LLM}} & \multicolumn{2}{c|}{\textbf{CES}} & \multicolumn{2}{c|}{\textbf{RAN}} & \multicolumn{2}{c|}{\textbf{SIM}} & \multicolumn{2}{c}{\textbf{LEN}} \\
         & \textit{Solo} & \textit{Unique} & \textit{Solo} & \textit{Unique} & \textit{Solo} & \textit{Unique} & \textit{Solo} & \textit{Unique} \\
        \midrule
        DeepSeek-R1 7B (Qwen) & 5.99\% & 2.24\% & 3.34\% & 0.06\% & 0.14\% & 0.03\% & 3.15\% & 2.25\% \\
        DeepSeek-R1 8B (Llama) & 7.26\% & 3.10\% & 3.69\% & 0.04\% & 0.09\% & 0.03\% & 2.49\% & 1.69\% \\
        Falcon-3 10B & 7.71\% & 4.07\% & 3.27\% & 0.00\% & 0.02\% & 0.01\% & 0.94\% & 0.45\% \\
        Falcon-3 7B & 7.68\% & 3.61\% & 3.64\% & 0.01\% & 0.02\% & 0.04\% & 1.24\% & 0.70\% \\
        GPT-OSS 120B & 8.32\% & 3.66\% & 4.41\% & 0.08\% & 0.02\% & 0.01\% & 0.70\% & 0.28\% \\
        GPT-OSS 20B & 8.72\% & 3.81\% & 4.59\% & 0.07\% & 0.02\% & 0.02\% & 0.96\% & 0.46\% \\
        Gemma-3 12B & 7.84\% & 3.41\% & 3.93\% & 0.04\% & 0.02\% & 0.07\% & 1.86\% & 1.24\% \\
        Gemma-3 27B & 7.46\% & 3.28\% & 3.74\% & 0.04\% & 0.04\% & 0.03\% & 1.75\% & 1.10\% \\
        Llama-3.1 8B & 7.44\% & 3.36\% & 3.65\% & 0.03\% & 0.01\% & 0.06\% & 1.32\% & 0.80\% \\
        \bottomrule
    \end{tabular}
    \vspace{5pt}
    \caption{Product feature contributions to different recommender LLMs decision rules ($R^2$). CES: Cross-Encoder Score; RAN: bi-encoder query--product cosine similarity. SIM: similarity to original description. LEN: description token length. Solo: marginal $R^2$ of each feature alone; Unique: $R^2$ after controlling for all other features.}
    \label{tab:variables_contribution}
\end{table}

% Observation 1: the length influences the decision rules, no matter the relevance (R2_unique)
Across all nine recommender models, description length ($\phi_{\text{LEN}}$) exhibits a positive unique contribution to the decision rule after controlling for all other features, with $R^2_{\text{unique, LEN}}$ ranging from $0.28\%$ (GPT-OSS~120B) to $2.25\%$ (DeepSeek-R1~7B Qwen) (see \autoref{tab:variables_contribution}). Crucially, this unique contribution persists even after accounting for query--product relevance signals ($\phi_{\text{CES}}$ and $\phi_{\text{RAN}}$) and semantic preservation ($\phi_{\text{SIM}}$), indicating that description length  influences the recommender's decisions beyond what the measured relevance and similarity signals explain.

% Observation 2: the models non-detected as biased present the less influece from length (GPT-OSS)
Notably, the two models for which the $\chi^2$ test did not reach statistical significance, GPT-OSS~120B and GPT-OSS~20B, also exhibit the smallest unique contributions of description length to their decision rules, with $R^2_{\text{unique, LEN}}$ of $0.28\%$ and $0.46\%$ respectively. While the absence of a statistically significant departure from uniformity does not imply the absence of structural bias (see Remark~\ref{rem:converse}), these results are consistent with a scenario in which neither semantic drift nor length exert a strong enough influence on these models to produce a detectable non-uniformity in paraphraser selection.

% Observation 3: the most contributive variable is the relevance
Across all models, query--product relevance as measured by $\phi_{\text{CES}}$ is the most contributive feature, with solo $R^2$ values ranging from $5.99\%$ (DeepSeek-R1~7B Qwen) to $8.72\%$ (GPT-OSS~20B) and unique $R^2$ values ranging from $2.24\%$ (DeepSeek-R1~7B Qwen) to $4.07\%$ (Falcon-3~10B). This is a sanity check result: a well-functioning recommender should primarily select products that are relevant to the user query, and the dominance of $\phi_{\text{CES}}$ over all surface-level features confirms that the models are not purely driven by superficial textual properties. While the absolute $R^2$ values may appear modest, this is expected when decomposing the decision rules of large-scale LLMs using smaller proxy models: the recommender's decision boundary is likely shaped by complex, high-dimensional patterns that smaller feature extractors such as \texttt{cross-encoder/ms-marco-MiniLM-L6-v2} can only partially capture, so the modest overall fit should not be interpreted as an absence of structure in the recommender's behavior.

% Observation 4: RAN feature overlaps
A second relevance proxy, $\phi_{\text{RAN}}$, exhibits a substantial solo $R^2$ ranging from $3.27\%$ (Falcon-3~10B) to $4.59\%$ (GPT-OSS~20B), yet its unique contribution collapses to near zero across all models, with values between $0.00\%$ (Falcon-3~10B) and $0.08\%$ (GPT-OSS~120B). This sharp discrepancy between solo and unique $R^2$ indicates that $\phi_{\text{RAN}}$ is largely redundant with $\phi_{\text{CES}}$: although they are computed by different models (a bi-encoder and a cross-encoder, respectively), both measure query--product relevance, and $\phi_{\text{CES}}$, which jointly encodes the query and the description and was fine-tuned specifically for passage ranking, effectively subsumes the relevance information carried by $\phi_{\text{RAN}}$ once both are included in the model.

% Observation 5: SIM does not influence the model
Finally, the semantic preservation proxy $\phi_{\text{SIM}}$, which measures the cosine similarity between original and paraphrased descriptions, exhibits negligible solo $R^2$ values across all models, ranging from $0.01\%$ (Llama-3.1~8B) to $0.14\%$ (DeepSeek-R1~7B Qwen), with unique contributions equally close to zero. This indicates that the degree to which a paraphraser preserves the original description has virtually no influence on the recommender's decisions, and that the recommender is primarily sensitive to how well a product description matches the user query rather than to how faithfully it reproduces the original wording. Combined with the dominance of $\phi_{\text{CES}}$ noted above, this suggests that among the measured features, query--product relevance is the principal semantic driver of the recommender's decision rule.

% Conclusion: length has orthogonal contribution, thus evidence of verbosity bias
Taken together, these observations provide evidence of verbosity bias in LLM-based RS. The dominance of $\phi_{\text{CES}}$ confirms that query--product relevance is the principal driver of the recommender's decision rule, a finding further supported by the redundancy of $\phi_{\text{RAN}}$ with $\phi_{\text{CES}}$ and the negligible contribution of $\phi_{\text{SIM}}$. Crucially, $\phi_{\text{LEN}}$ retains a modest unique $R^2$ after controlling for all relevance and semantic preservation signals, confirming that description length adds explanatory power beyond what the measured relevance signals provide. The direction and magnitude of this effect, as well as the interaction between verbosity and relevance when they provide conflicting signals, are further examined and re-confirmed in the following steps.

\subsubsection{Variables direction}

% Summary on what is this step about and present results (reference table, one sentence)
This step examines the direction of each feature's effect on the recommender's decision rule, determining whether higher values of a given feature increase or decrease the probability of a product being selected. \autoref{tab:variables_direction} reports the solo and unique coefficients of the conditional logit model for all four features across all nine recommender models, revealing that description length has a consistently positive effect on product selection that persists after controlling for all relevance and semantic preservation signals.

\begin{table}
    \centering
    \begin{tabular}{l|c|c|c|c|c|r}
    \toprule
    \multirow{2}{*}{\textbf{Recommender LLM}} & \multirow{2}{*}{$\beta^{\text{solo}}_{\text{len}}$} & \multicolumn{4}{c|}{$\textbf{Coefficients}$} & \multirow{2}{*}{\textbf{Atten.}} \\
     & & $\beta^{\text{unique}}_{\text{CES}}$ & $\beta^{\text{unique}}_{\text{RAN}}$ & $\beta^{\text{unique}}_{\text{SIM}}$ & $\beta^{\text{unique}}_{\text{LEN}}$ \\
    \midrule
    DeepSeek-R1 7B (Qwen) & \cellcolor{posgreen}+0.3505 & \cellcolor{posgreen}+0.5461 & \cellcolor{posgreen}+0.0975 & \cellcolor{negred}-0.0407 & \cellcolor{posgreen}+0.3220 & 8.15\% \\
    DeepSeek-R1 8B (Llama) & \cellcolor{posgreen}+0.3119 & \cellcolor{posgreen}+0.6471 & \cellcolor{posgreen}+0.0755 & \cellcolor{negred}-0.0402 & \cellcolor{posgreen}+0.2811 & 9.87\% \\
    Falcon-3 10B & \cellcolor{posgreen}+0.1930 & \cellcolor{posgreen}+0.7373 & \cellcolor{negred}-0.0102 & \cellcolor{negred}-0.0231 & \cellcolor{posgreen}+0.1469 & 23.90\% \\
    Falcon-3 7B & \cellcolor{posgreen}+0.2221 & \cellcolor{posgreen}+0.6939 & \cellcolor{posgreen}+0.0436 & \cellcolor{negred}-0.0427 & \cellcolor{posgreen}+0.1826 & 17.79\% \\
    GPT-OSS 120B & \cellcolor{posgreen}+0.1672 & \cellcolor{posgreen}+0.7000 & \cellcolor{posgreen}+0.1065 & \cellcolor{negred}-0.0245 & \cellcolor{posgreen}+0.1167 & 30.17\% \\
    GPT-OSS 20B & \cellcolor{posgreen}+0.1957 & \cellcolor{posgreen}+0.7174 & \cellcolor{posgreen}+0.1052 & \cellcolor{negred}-0.0291 & \cellcolor{posgreen}+0.1484 & 24.20\% \\
    Gemma-3 12B & \cellcolor{posgreen}+0.2706 & \cellcolor{posgreen}+0.6784 & \cellcolor{posgreen}+0.0799 & \cellcolor{negred}-0.0616 & \cellcolor{posgreen}+0.2423 & 10.44\% \\
    Gemma-3 27B & \cellcolor{posgreen}+0.2621 & \cellcolor{posgreen}+0.6639 & \cellcolor{posgreen}+0.0729 & \cellcolor{negred}-0.0401 & \cellcolor{posgreen}+0.2271 & 13.35\% \\
    Llama-3.1 8B & \cellcolor{posgreen}+0.2283 & \cellcolor{posgreen}+0.6696 & \cellcolor{posgreen}+0.0666 & \cellcolor{negred}-0.0530 & \cellcolor{posgreen}+0.1944 & 14.82\% \\
    \bottomrule
\end{tabular}
    \vspace{5pt}
    \caption{$\beta$ coefficients from a CLogit predicting paraphraser selection. A positive (negative) value indicates that higher values of the variable increase (decrease) the probability of the product being selected by the recommender LLM. $\beta^{\text{solo}}_{\text{len}}$: length coefficient estimated without controls. Atten.: percentage reduction in $\beta_{\text{len}}$ when all controls are included.}
    \label{tab:variables_direction}
\end{table}

% Observation 1: length contribution is positive
The solo coefficient of description length $\hat{\beta}^{\text{solo}}_{\text{LEN}}$ is positive across all nine recommender models, ranging from $+0.1672$ (GPT-OSS 120B) to $+0.3505$ (DeepSeek-R1~7B Qwen), indicating that longer product descriptions consistently increase the probability of being selected by the recommender. Importantly, this positive effect persists after controlling for all other features, with $\hat{\beta}^{\text{unique}}_{\text{LEN}}$ remaining positive across all models, ranging from $+0.1167$ (GPT-OSS~120B) to $+0.3220$ (DeepSeek-R1~7B Qwen). This shows that the positive effect of description length is robust to controlling for the measured relevance and semantic similarity signals. Since all features are standardized, $e^{\hat{\beta}}$ can be read as the multiplicative change in the odds of selection per standard deviation of the feature: for instance, $\hat{\beta}^{\text{unique}}_{\text{LEN}} = 0.322$ for DeepSeek-R1~7B (Qwen) corresponds to a $38\%$ increase in the odds of selection per standard deviation of description length, and $\hat{\beta}^{\text{unique}}_{\text{LEN}} = 0.117$ for GPT-OSS~120B to a $12\%$ increase.

% Observation 2: attenuation is higher for less biased models
The attenuation of $\hat{\beta}_{\text{LEN}}$ when controlling for all other features is consistently higher for models that showed weaker or non-significant bias in the detection step, with GPT-OSS~120B and GPT-OSS~20B exhibiting the largest reductions of $30.17\%$ and $24.20\%$ respectively, compared to models such as DeepSeek-R1~7B (Qwen) and DeepSeek-R1~8B (Llama), which show attenuations of only $8.15\%$ and $9.87\%$. This suggests that in models where verbosity bias is weaker, a larger fraction of the raw length effect is explained away by relevance and semantic features, whereas in strongly biased models the length coefficient remains largely stable under control, pointing to a more deeply ingrained sensitivity to description length.

% Observation 3: relevance is positive (sanity check)
The unique coefficient of query--product relevance $\hat{\beta}^{\text{unique}}_{\text{CES}}$ is positive across all nine recommender models, ranging from $+0.5461$ (DeepSeek-R1~7B Qwen) to $+0.7373$ (Falcon-3~10B), confirming that higher relevance consistently increases the probability of a product being selected. This is a sanity check result: a well-functioning recommender should favor products that better match the user query, and the uniformly positive and largest coefficients of $\phi_{\text{CES}}$ across all models confirms that relevance remains the dominant and expected driver of the recommender's decision rule, lending credibility to the independent length effect identified alongside it.

% Observation 4: similarity is negative but very small
The unique coefficient of semantic preservation $\hat{\beta}^{\text{unique}}_{\text{SIM}}$ is negative across all nine recommender models, ranging from $-0.0231$ (Falcon-3~10B) to $-0.0616$ (Gemma-3~12B), indicating that higher similarity to the original description slightly decreases the probability of selection. While this effect is very small in magnitude, its negative direction is consistent with the selection patterns observed in \autoref{tab:uniformity_test}: the identity paraphraser, which always produces a similarity of $1$ by definition, is among the least preferred paraphrasers across most models, naturally pulling the coefficient in the negative direction. This should not be interpreted as recommenders penalizing semantic faithfulness per se, but rather as a reflection of the systematic underselection of the identity paraphraser driving the estimated coefficient.

% Conclusion: verbosity is positive
Taken together, the direction of the estimated coefficients provides a coherent and consistent picture across all nine recommender models. Relevance, as measured by $\phi_{\text{CES}}$, is the dominant positive driver of selection, confirming that the recommender behaves as expected in its primary function. Crucially, description length exhibits a robust positive effect that persists after controlling for all other features, directly establishing the direction of verbosity bias: recommender models systematically favor longer product descriptions beyond what the measured query--product relevance signals explain. This directional finding complements the $R^2$ decomposition of the previous step, and is further examined in the next section, where the interaction between verbosity and relevance when they provide conflicting signals is explored in detail.

\subsubsection{Conditional analysis of conflicting signals}

\begin{figure}[t]
    \begin{subfigure}[t]{\linewidth}
        \includegraphics[width=\linewidth]{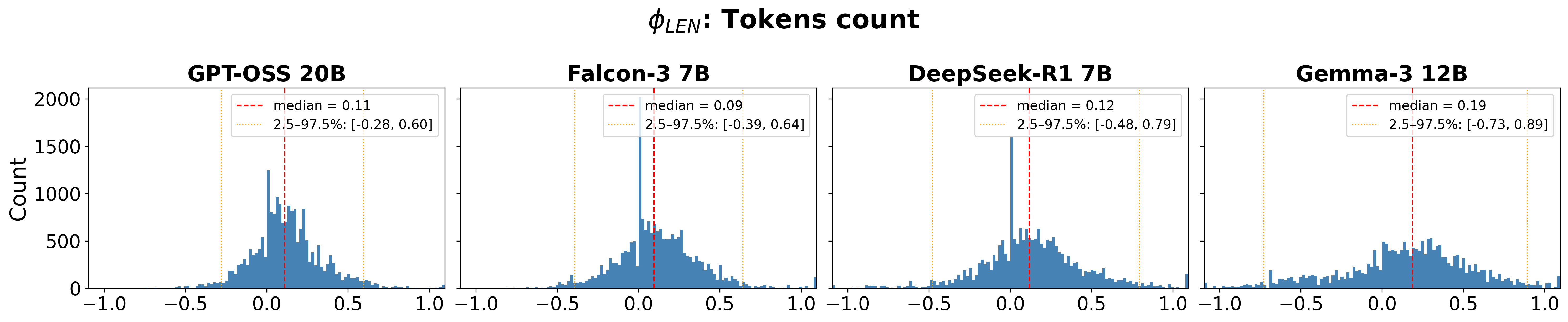}
    \end{subfigure}
    \begin{subfigure}[t]{\linewidth}
        \includegraphics[width=\linewidth]{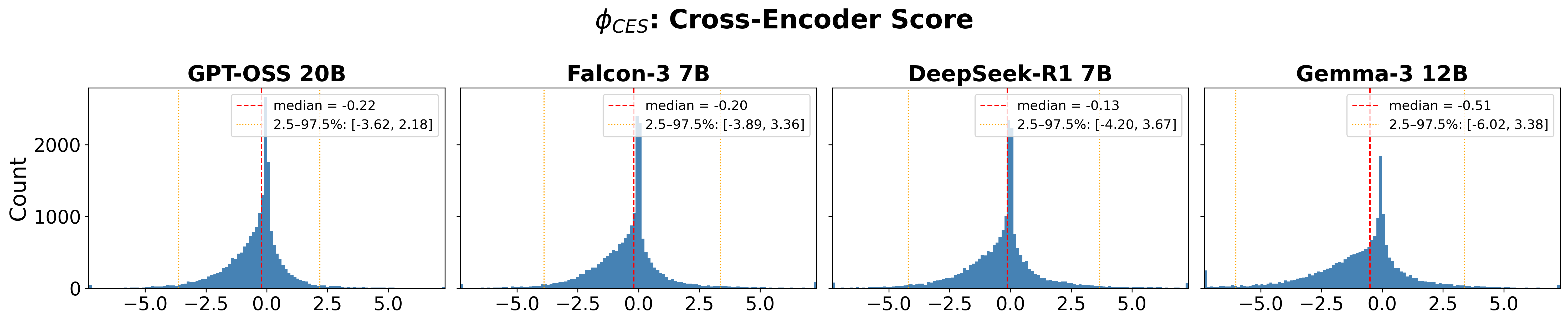}
    \end{subfigure}
    \caption{Histograms of the feature differences between each original product description and its paraphrase, one panel per LLM paraphraser $\tau_c$. For each query $q$ and each candidate product $x$ rewritten by $\tau_c$, the original description $x$ is compared with its paraphrase $\tau_c(x)$ through $\Delta_{\mathrm{LEN}} = \log\left(\phi_{\mathrm{LEN}}(x) / \phi_{\mathrm{LEN}}(\tau_c(x))\right)$ (top) and $\Delta_{\mathrm{CES}} = \phi_{\mathrm{CES}}(q, x) - \phi_{\mathrm{CES}}(q, \tau_c(x))$ (bottom), using the same log-scale differences as Step~4. Positive values mean that the original description is longer or more relevant than its paraphrase. Red dashed lines mark the median, orange dotted lines the 2.5th and 97.5th percentiles.}
    \label{fig:histograms}
\end{figure}

% Summary on what is this step about. First justification and then present results
This step applies Step~4 of \autoref{sec:framework:methodology} with $\phi_{\text{LEN}}$ as the surface-level feature and $\phi_{\text{CES}}$ as the semantic feature, taking the identity paraphraser as the focal paraphraser and each of the four LLM paraphrasers as the comparison. The reported gaps therefore measure how the recommender's preference for the original description changes depending on whether it is longer or shorter, and more or less relevant, than the paraphrased description it is compared with. Length is compared through the log-ratio $\Delta_{\mathrm{LEN}} = \log\left(\phi_{\mathrm{LEN}}(x_{q,f}) / \phi_{\mathrm{LEN}}(x_{q,c})\right)$ and relevance through the difference of cross-encoder logits $\Delta_{\mathrm{CES}} = \phi_{\mathrm{CES}}(q, x_{q,f}) - \phi_{\mathrm{CES}}(q, x_{q,c})$, i.e.\ the log-ratio of the odds of relevance, so the $=$ label means that the two lengths, or the two odds of relevance, differ by less than a factor of $1.1$. Of the pairs used in this step, only $6.6$--$7.4\%$ fall inside the $=$ band for $\phi_{\text{LEN}}$ and $3.5$--$4.0\%$ for $\phi_{\text{CES}}$, depending on the comparison paraphraser, so the threshold isolates pairs with near-identical values while leaving most pairs in the directional conditions. Since these pairs involve two different products, their differences combine the variation between products with the effect of the paraphraser. \autoref{fig:histograms} isolates the latter by comparing each description with its own paraphrase: both differences are peaked at zero with long tails, and only 18–35\% of the paraphrases fall within the length tolerance and 12–21\% within the CES tolerance relative to the original description. The paraphrasers alone therefore move most descriptions outside the $=$ band. The results are reported in \autoref{tab:variables_analysis}, and the following observations characterize the key patterns emerging from the condition stratification.

% Observation 1: all conditions generates a minimum number of samples big enough
All eight joint conditions are populated for every recommender model. As reported in the minimum samples block of \autoref{tab:variables_analysis}, the least populated condition is $(-,=)$, with a minimum of 191 samples (DeepSeek-R1~7B Qwen), while the aligned conditions $(+,+)$ and $(-,-)$ exceed 4{,}100 samples in every model. The one-signal-clear conditions are therefore estimated with less precision than the others, so we interpret them through the consistency of their sign across models rather than through the value obtained for any single model.

\begin{table}
\begin{tabular}{c|l|r|r|r|r|r|r|r|r}
    \toprule
    & \multirow{2}{*}{\textbf{Recommender LLM}} & \multicolumn{4}{c|}{\textbf{Both signals clear}} & \multicolumn{4}{c}{\textbf{One signal clear}} \\
    & & \multicolumn{1}{c|}{$({+}{+})$} & \multicolumn{1}{c|}{$({+}{-})$} & \multicolumn{1}{c|}{$({-}{+})$} & \multicolumn{1}{c|}{$({-}{-})$} & \multicolumn{1}{c|}{$({+}{=})$} & \multicolumn{1}{c|}{$({=}{+})$} & \multicolumn{1}{c|}{$({=}{-})$} & \multicolumn{1}{c}{$({-}{=})$} \\
    \midrule
    \multirow{9}{*}{\rotatebox{90}{\textit{Selection gap}}} & DeepSeek-R1 7B (Qwen) & \cellcolor{posgreen}+20.2\% & \cellcolor{negred}-0.3\% & \cellcolor{posgreen}+2.4\% & \cellcolor{negred}-16.8\% & \cellcolor{posgreen}+4.9\% & \cellcolor{posgreen}+10.9\% & \cellcolor{negred}-12.1\% & \cellcolor{negred}-1.9\% \\
     & DeepSeek-R1 8B (Llama) & \cellcolor{posgreen}+18.1\% & \cellcolor{negred}-4.0\% & \cellcolor{posgreen}+4.0\% & \cellcolor{negred}-18.0\% & \cellcolor{posgreen}+2.7\% & \cellcolor{posgreen}+10.6\% & \cellcolor{negred}-12.8\% & \cellcolor{negred}-1.3\% \\
     & Falcon-3 10B & \cellcolor{posgreen}+13.8\% & \cellcolor{negred}-8.4\% & \cellcolor{posgreen}+6.4\% & \cellcolor{negred}-15.8\% & \cellcolor{posgreen}+2.8\% & \cellcolor{posgreen}+12.3\% & \cellcolor{negred}-13.4\% & \cellcolor{negred}-0.8\% \\
     & Falcon-3 7B & \cellcolor{posgreen}+15.2\% & \cellcolor{negred}-7.5\% & \cellcolor{posgreen}+6.0\% & \cellcolor{negred}-17.1\% & \cellcolor{posgreen}+2.2\% & \cellcolor{posgreen}+12.8\% & \cellcolor{negred}-12.5\% & \cellcolor{negred}-1.8\% \\
     & GPT-OSS 120B & \cellcolor{posgreen}+14.8\% & \cellcolor{negred}-9.9\% & \cellcolor{posgreen}+10.1\% & \cellcolor{negred}-14.3\% & \cellcolor{posgreen}+0.9\% & \cellcolor{posgreen}+12.3\% & \cellcolor{negred}-13.0\% & \cellcolor{posgreen}+1.9\% \\
     & GPT-OSS 20B & \cellcolor{posgreen}+17.0\% & \cellcolor{negred}-8.8\% & \cellcolor{posgreen}+10.7\% & \cellcolor{negred}-14.8\% & \cellcolor{posgreen}+0.3\% & \cellcolor{posgreen}+13.1\% & \cellcolor{negred}-14.1\% & \cellcolor{negred}-0.6\% \\
     & Gemma-3 12B & \cellcolor{posgreen}+16.6\% & \cellcolor{negred}-6.5\% & \cellcolor{posgreen}+6.5\% & \cellcolor{negred}-17.2\% & \cellcolor{posgreen}+3.7\% & \cellcolor{posgreen}+11.4\% & \cellcolor{negred}-14.9\% & \cellcolor{negred}-1.8\% \\
     & Gemma-3 27B & \cellcolor{posgreen}+16.4\% & \cellcolor{negred}-6.0\% & \cellcolor{posgreen}+6.4\% & \cellcolor{negred}-16.9\% & \cellcolor{posgreen}+2.3\% & \cellcolor{posgreen}+11.1\% & \cellcolor{negred}-12.4\% & \cellcolor{posgreen}+0.6\% \\
     & Llama-3.1 8B & \cellcolor{posgreen}+14.9\% & \cellcolor{negred}-7.0\% & \cellcolor{posgreen}+6.4\% & \cellcolor{negred}-16.2\% & \cellcolor{posgreen}+0.6\% & \cellcolor{posgreen}+7.8\% & \cellcolor{negred}-13.1\% & \cellcolor{negred}-0.7\% \\
    \midrule
    \multirow{9}{*}{\rotatebox{90}{\textit{Min. samples}}} & DeepSeek-R1 7B (Qwen) & 4473 & 4392 & 2311 & 4108 & 281 & 380 & 677 & 191 \\
     & DeepSeek-R1 8B (Llama) & 4947 & 4794 & 2543 & 4535 & 364 & 431 & 734 & 236 \\
     & Falcon-3 10B & 4805 & 4678 & 2467 & 4431 & 352 & 424 & 717 & 235 \\
     & Falcon-3 7B & 4788 & 4651 & 2478 & 4403 & 341 & 414 & 716 & 224 \\
     & GPT-OSS 120B & 4969 & 4811 & 2556 & 4558 & 370 & 432 & 736 & 238 \\
     & GPT-OSS 20B & 4904 & 4763 & 2537 & 4522 & 362 & 428 & 733 & 233 \\
     & Gemma-3 12B & 4933 & 4788 & 2544 & 4537 & 363 & 428 & 739 & 233 \\
     & Gemma-3 27B & 4971 & 4810 & 2565 & 4555 & 369 & 430 & 734 & 238 \\
     & Llama-3.1 8B & 4928 & 4774 & 2538 & 4521 & 364 & 430 & 733 & 235 \\
    \bottomrule
\end{tabular}
    \vspace{5pt}
    \caption{Average selection probability gap (\autoref{eq:conditional_gap_avg}) between the focal paraphraser (identity) and the four comparison paraphrasers, stratified by joint LEN--CES conditions. Each condition is a pair in $\{+,-,=\}^2$: the first coordinate refers to description length ($\phi_{\mathrm{LEN}}$) and the second to the cross-encoder score ($\phi_{\mathrm{CES}}$). For each query $q$ and comparison paraphraser $c$, a $+$ ($-$) indicates that the difference $\Delta$ between the original description $x_{q,f}$ and the paraphrased description $x_{q,c}$ is larger than $\log(1.1)$ (smaller than $-\log(1.1)$), where $\Delta$ is the log-ratio of lengths for $\phi_{\mathrm{LEN}}$ and the difference of cross-encoder logits for $\phi_{\mathrm{CES}}$; an $=$ indicates that $|\Delta| \le \log(1.1)$. Green cells indicate that the original description is preferred by the recommender, and red cells that it is avoided. The minimum samples block reports, for each recommender model and condition, the smallest number of queries satisfying the condition across the four comparison paraphrasers.}
    \label{tab:variables_analysis}
\end{table}

% Observation 2: for both signal clear and same direction (++, --), the gap between the focal model and the comp model is high
When both signals are clear and aligned, the selection gaps are large and consistently signed across all nine recommender models. Under condition $(+,+)$, where the focal paraphraser produces a longer description \emph{and} a higher relevance score than the comparison, the average selection gap ranges from $+13.8\%$ (Falcon-3~10B) to $+20.2\%$ (DeepSeek-R1~7B Qwen), confirming that the focal paraphraser under this condition is strongly preferred. Symmetrically, under condition $(-,-)$, where the focal paraphraser is both shorter and less relevant, the gap ranges from $-14.3\%$ (GPT-OSS~120B) to $-18.0\%$ (DeepSeek-R1~8B Llama), confirming that the focal paraphraser is strongly avoided. The consistency of these large, correctly-signed gaps across all models serves as a sanity check for the methodology: when both verbosity and relevance point in the same direction, the recommender responds in the expected direction with substantial magnitude, validating that the condition stratification is capturing genuine differences in product descriptions as perceived by the recommender.

% Observation 3: for both signal clear, relevance is the most important variable, but the gaps are significantly lower (indicates both variables contribute).
When both signals are clear but point in opposite directions, the selection gaps reveal that relevance dominates over verbosity when the two features conflict, though the gaps are substantially attenuated compared to the aligned conditions. Under condition $(+,-)$, where the focal paraphraser is longer but less relevant, the gap is negative across all models, ranging from $-0.3\%$ (DeepSeek-R1~7B Qwen) to $-9.9\%$ (GPT-OSS~120B), indicating that the relevance disadvantage overrides the length advantage in driving the recommender's decision. Under condition $(-,+)$, where the focal paraphraser is shorter but more relevant, the gap is positive across all models, ranging from $+2.4\%$ (DeepSeek-R1~7B Qwen) to $+10.7\%$ (GPT-OSS~20B), confirming that a relevance advantage is sufficient to overcome a length disadvantage. Crucially, the magnitude of these conflicting-signal gaps is considerably smaller than those observed under aligned conditions: the absolute gap under $(+,-)$ and $(-,+)$ averages $6.5\%$, against $16.3\%$ under $(+,+)$ and $(-,-)$, indicating that while relevance ultimately prevails, the opposing verbosity signal partially counteracts it, providing direct empirical evidence that both features contribute to the recommender's decision rule.

% Observation 4: when a signal is clear, while relevance has the higher gap, for equal relevance the verbosity plays a role.
The one-signal-clear conditions isolate the marginal effect of each feature in the absence of interference from the other, providing the cleanest estimate of their independent contributions. Under condition $(=,+)$, where the focal paraphraser has similar length but higher relevance, the selection gap is positive across all models, ranging from $+7.8\%$ (Llama-3.1~8B) to $+13.1\%$ (GPT-OSS~20B), confirming that relevance alone is sufficient to drive a substantial preference for the focal paraphraser. Under condition $(=,-)$, the gap is symmetrically negative, ranging from $-12.1\%$ (DeepSeek-R1~7B Qwen) to $-14.9\%$ (Gemma-3~12B), confirming the symmetric penalty for a relevance disadvantage when length is held equal. By contrast, under condition $(+,=)$, where the focal paraphraser is longer but has similar relevance, the gap is considerably smaller but positive for all nine models, ranging from $+0.3\%$ (GPT-OSS~20B) to $+4.9\%$ (DeepSeek-R1~7B Qwen), and under condition $(-,=)$, where the focal paraphraser is shorter with similar relevance, the gap is negative for seven of the nine models (the exceptions being GPT-OSS~120B, $+1.9\%$, and Gemma-3~27B, $+0.6\%$) and equally modest, reaching at most $-1.9\%$ (DeepSeek-R1~7B Qwen). These results confirm the hierarchy established in the previous steps: relevance is the primary driver of selection, producing gaps several times larger than those driven by verbosity alone (on average $12.3\%$ against $1.8\%$ in absolute value), yet verbosity retains a consistent and correctly-signed independent effect across almost all models even when relevance provides no differential signal.

% Observation 5: for GPT-OSS model (the ones not detected as biased) the verbosity has almost no weight in the decision process (this is specially observed in one signal clear), but verbosity helps the decision.
The GPT-OSS models, which were the only two recommenders for which the $\chi^2$ test did not reach statistical significance in Step~1, show the weakest response when verbosity is the only differential signal. Under condition $(+,=)$, GPT-OSS~120B and GPT-OSS~20B produce gaps of $+0.9\%$ and $+0.3\%$, the smallest together with Llama-3.1~8B ($+0.6\%$), whereas DeepSeek-R1~7B (Qwen) reaches $+4.9\%$. Under condition $(-,=)$, the gaps of all models are small, between $-1.9\%$ and $+1.9\%$, and the two positive values (GPT-OSS~120B and Gemma-3~27B) may reflect the limited size of this condition, so we do not interpret them individually. This weak response to isolated verbosity signals is consistent with the smallest unique $R^2_{\text{LEN}}$ and largest attenuation values observed for these models in Steps~2 and~3, and collectively suggests that the GPT-OSS family is substantially less sensitive to description length as a standalone cue. Consistently, their conflicting-signal gaps under $(+,-)$ and $(-,+)$ are the largest observed ($-9.9\%$ and $-8.8\%$, and $+10.1\%$ and $+10.7\%$, respectively), meaning that a relevance advantage is the least eroded by an opposing length difference in these models. Their gaps under $(+,+)$ and $(-,-)$ are larger in magnitude than under $(=,+)$ and $(=,-)$, which may suggest that length reinforces a concurrent relevance signal; however, these conditions also differ in the distribution of relevance differences, so this comparison does not isolate an interaction effect and we do not draw conclusions from it. Finally, both GPT-OSS models have a positive unique length coefficient in Step~3, so their sensitivity to description length is weak rather than absent.

% Conclusions
Taken together, the conditional analysis of conflicting signals provides a fine-grained and coherent characterization of how verbosity and relevance jointly drive product selection across all nine recommender models. The aligned conditions $(+,+)$ and $(-,-)$ confirm that when both signals point in the same direction the recommender responds with large and consistent gaps, validating the condition stratification as a reliable probe of the recommender's decision rule. The conflicting conditions $(+,-)$ and $(-,+)$ establish a clear hierarchy: relevance dominates over verbosity when the two features disagree, yet the substantial attenuation of the gaps relative to the aligned conditions reveals that verbosity is not merely noise but an independent signal that partially counteracts the relevance advantage. The one-signal-clear conditions $(+,=)$ and $(-,=)$ provide the most direct evidence of verbosity bias in isolation, showing that, in the absence of a differential relevance signal, a longer original description is preferred in all nine models and a shorter one is avoided in seven, although these gaps are small. The GPT-OSS models lie at the weak end of this spectrum of verbosity sensitivity: their response to isolated length differences is small but positive, consistent with a verbosity bias that is present but weak. Across all conditions and all models, these findings re-confirm and extend the conclusions of Steps~1--3: relevance is the primary driver of the recommender's decision rule, but description length operates as a robust, independent, and consistently positive secondary signal whose influence is detectable both in isolation and in interaction with relevance.

\section{Conclusions and Future Work}
\label{sec:conclusions}

% --- Findings -------------------------------------------------------------
This paper introduced a label-free, statistically grounded methodology for detecting structural biases in LLM-based RS and applied it to verbosity bias, i.e., the tendency to prefer items with longer descriptions, across nine recommender LLMs from five model families. The methodology rests on a simple guarantee: if the recommender is insensitive to surface-level features and the paraphrasers preserve meaning, the selected paraphraser must be uniformly distributed (Proposition~\ref{prop:uniformity_proposition}). Any departure from uniformity is therefore a measurable signal, and it can be obtained without any relevance annotation.

% How we solve RQ1-3. Verbosity has genuine contribution to decision rules
Our three research questions concerning verbosity bias in LLM-based RS receive clear answers. \textbf{RQ1} asked whether LLM-based RS exhibit verbosity bias in product selection: seven of the nine recommenders select paraphrased descriptions non-uniformly, showing that semantically equivalent rewritings of the same products are not treated as interchangeable, and the subsequent steps of the methodology indicate that description length is one of the factors behind this non-uniformity. \textbf{RQ2} asked whether verbosity independently influences product selection beyond query--product relevance signals. The answer is affirmative: the conditional logit analysis of Steps~2 and~3 shows that description length retains a non-negligible unique $R^2$ and a positive unique coefficient in all nine models after controlling for all relevance and semantic similarity signals, so longer descriptions are favored beyond what the measured relevance signals explain. \textbf{RQ3} asked how verbosity and relevance interact when they provide conflicting signals. The conditional analysis of conflicting signals of Step~4 shows that relevance remains the primary driver of the recommender's decisions: when length and relevance point in opposite directions, the more relevant description is still preferred. However, the preference gap in these conflicting cases is on average less than half that observed when both signals agree, indicating that verbosity acts as a secondary signal that partially counteracts relevance rather than as noise. Moreover, when length is the only differing signal, the longer description is preferred by all nine recommenders, although by a small margin.

The overall picture is therefore reassuring and cautionary at the same time. The evaluated recommenders behave as recommenders should, in that relevance dominates their choices. Yet all of them also respond, to different degrees, to how much is written about a product. The magnitude of this effect is model-dependent. The GPT-OSS models are the least sensitive, with the smallest unique length effects in Steps~2--3 and among the smallest responses to isolated length differences in Step~4. The DeepSeek-R1 distillations are the most sensitive. This indicates that verbosity sensitivity is not an inevitable property of LLM-based recommendation but varies with model design and training. This is consistent with broader evidence that reliability properties of LLMs, including stability to semantically equivalent rephrasings of the prompt, do not improve uniformly with scale or instruction tuning and differ across model families~\cite{Zhou2024}. Such properties must therefore be audited empirically for each deployed model rather than assumed.

% --- Implications ---------------------------------------------------------

These findings have direct implications for the different stakeholders of LLM-based recommendation.

\textit{Platform operators.} Verbosity sensitivity varies by a factor of almost three across the evaluated models (unique length coefficients from $0.12$ to $0.32$), so it should be part of model selection alongside accuracy and cost. Because the audit of Algorithm~\ref{alg:audit} needs only black-box access, reusable paraphrases and a few thousand queries, it can be integrated into the validation pipeline and rerun whenever the model, the prompt template or the catalog changes, in the same way as regression tests in software deployment.

\textit{Marketplaces and sellers.} When descriptions are written by sellers, a recommender that rewards longer descriptions creates an incentive to pad listings regardless of product quality, and disadvantages sellers who describe their products concisely. Detecting this sensitivity is a prerequisite for ensuring that recommendations reflect product fit rather than writing style, and the audit provides quantitative evidence for platform transparency and risk-assessment obligations~\cite{dsa2022}.

\textit{System designers.} The results point to inexpensive mitigations that can be evaluated with the same framework: normalizing or templating descriptions before they reach the recommender (for instance, truncating them to a common length or rewriting all candidates with the same paraphraser), preferring the less length-sensitive models when descriptions vary widely in length, and adapting post-hoc calibration techniques developed for length bias in LLM evaluation~\cite{dubois2024length, huang2025posthoc}. Because the audit outputs statistically testable measures, it can confirm whether a mitigation actually reduces the bias rather than merely changing the recommendations.
 
% --- Future work ----------------------------------------------------------
Our empirical findings concern single-turn product recommendation with synthetic user queries, $k=5$ candidates and deterministic decoding, and Steps~2--4 estimate associations between description features and selection rather than causal effects. The methodology itself, however, is agnostic to the recommendation domain, the number of candidates and paraphrasers, and the features under study: it applies to any recommender that selects among textual candidates. This generality opens several directions for future work.
\begin{itemize}
    \item \textbf{Beyond verbosity.} The framework can be instantiated on other surface-level properties, such as lexical choices (e.g.\ adjective density), sentiment, formatting, or stylistic markers of LLM-generated text. This would give a more complete map of the structural biases that shape LLM-based recommendation.
    \item \textbf{Richer recommender architectures.} The audit can be extended to recommenders that enrich the candidate list with LLM-generated descriptions, collaborative signals or reasoning LLMs, such as RALLRec+~\cite{LUO2026129508}. These designs change both the text the recommender reads and how it decides, so they may amplify or reduce structural biases. This question is especially relevant given that the reasoning-distilled DeepSeek-R1 models were the most length-sensitive recommenders in our study.
    \item \textbf{Mitigation benchmarking.} Because the framework yields quantitative, statistically testable measures of bias, it can serve as an evaluation protocol for mitigation strategies, from description normalization to bias-aware prompting and post-hoc recalibration. A natural first candidate is the approach of \cite{huang2025posthoc}, which estimates the length-dependent component of reward-model scores and removes it post hoc; adapting it to the selection scores of LLM-based recommenders and measuring its effect with our framework is a promising direction.
    \item \textbf{Measuring the cost to users.} Combining the label-free analysis with human relevance judgments or interaction data would quantify how often verbosity bias changes the recommendation to a less relevant item, and thereby measure its practical cost to users.
    \item \textbf{Verbosity-agnostic relevance signals.} Any additional information carried by a longer description should be reflected both in the recommender's choice and in the cross-encoder relevance score, so a preference for longer descriptions that remains after controlling for relevance is most plausibly explained by verbosity bias. This reasoning relies on the relevance signal itself being insensitive to length. Future work could build relevance signals that are agnostic to verbosity by construction, for instance by decomposing each description into atomic facts, removing redundant ones, and computing relevance from the resulting set of facts, so that restating or padding the same content leaves the score unchanged. The framework proposed in this paper allows new features of this kind to be analyzed automatically.
\end{itemize}
 
Overall, this work shows that structural biases in LLM-based recommender systems can be detected and characterized rigorously without relevance labels. Verbosity is one bias that is now measurable, and the same framework is ready to be applied to others. Since it requires neither annotations nor access to model internals, it can become a standard, low-cost step in the validation and monitoring of deployed LLM-based recommender systems.

\subsubsection*{Acknowledgments} This publication is part of the Project “Ethical, Responsible and General Purpose Artificial Intelligence: Applications In Risk Scenarios” (IAFER) Exp.:TSI-100927-2023-1 funded through the Creation of university-industry research programs (Enia Programs), aimed at the research and development of artificial intelligence, for its dissemination and education within the framework of the Recovery, Transformation and Resilience Plan from the European Union Next Generation EU through the Ministry for Digital Transformation and the Civil Service.

\subsection*{Declaration of generative AI and AI-assisted technologies in the writing process}
During the preparation of this work the authors used Claude (Anthropic) in order to improve the language of the manuscript and to review its statistical arguments. After using this tool, the authors reviewed and edited the content as needed and take full responsibility for the content of the publication.

\bibliographystyle{splncs04}
\bibliography{biblatex}

\end{document}